\input harvmac.tex


\let\includefigures=\iftrue
\let\useblackboard=\iftrue
\newfam\black

\includefigures
\message{If you do not have epsf.tex (to include figures),}
\message{change the option at the top of the tex file.}
\input epsf
\def\figin{\epsfcheck\figin}\def\figins{\epsfcheck\figins}
\def\epsfcheck{\ifx\epsfbox\UnDeFiNeD
\message{(NO epsf.tex, FIGURES WILL BE IGNORED)}
\gdef\figin##1{\vskip2in}\gdef\figins##1{\hskip.5in}
\else\message{(FIGURES WILL BE INCLUDED)}%
\gdef\figin##1{##1}\gdef\figins##1{##1}\fi}
\def\DefWarn#1{}
\def\figinsert{\goodbreak\midinsert}
\def\ifig#1#2#3{\DefWarn#1\xdef#1{fig.~\the\figno}
\writedef{#1\leftbracket fig.\noexpand~\the\figno}%
\figinsert\figin{\centerline{#3}}\medskip\centerline{\vbox{
\baselineskip12pt\advance\hsize by -1truein
\noindent\footnotefont{\bf Fig.~\the\figno:} #2}}
\endinsert\global\advance\figno by1}
\else
\def\ifig#1#2#3{\xdef#1{fig.~\the\figno}
\writedef{#1\leftbracket fig.\noexpand~\the\figno}%
\global\advance\figno by1} \fi

\def \la {\langle}
\def \ra {\rangle}

\def \pa {\partial}

\def \eps {\epsilon}
\def\vev#1{\left\langle #1 \right\rangle}
\def\OO{{\cal OO}}

\catcode`\@=11
\def\slash#1{\mathord{\mathpalette\c@ncel{#1}}}
\overfullrule=0pt
\def\AA{{\cal A}}

\def\CC{{\cal C}}

\def\EE{{\cal E}}

\def\II{{\cal I}}

\def\OO{{\cal O}}

\def\TT{{\cal T}}

\def\II{{\cal I}}

\def\eps{\epsilon}

\def\underrel#1\over#2{\mathrel{\mathop{\kern\z@#1}\limits_{#2}}}

\catcode`\@=12


\def\ket#1{\left| #1\right\rangle}
\def\vev#1{\left\langle #1 \right\rangle}

\def\exp{{\rm exp}}


\def\p{{\partial}}


\lref\HeemskerkPN{
  I.~Heemskerk, J.~Penedones, J.~Polchinski and J.~Sully,
  ``Holography from Conformal Field Theory,''
JHEP {\bf 0910}, 079 (2009).
[arXiv:0907.0151 [hep-th]].
}

\lref\GrinsteinQK{
  B.~Grinstein, K.~A.~Intriligator and I.~Z.~Rothstein,
  ``Comments on Unparticles,''
Phys.\ Lett.\ B {\bf 662}, 367 (2008).
[arXiv:0801.1140 [hep-ph]].
}

\lref\CornalbaXK{
  L.~Cornalba, M.~S.~Costa, J.~Penedones and R.~Schiappa,
  ``Eikonal Approximation in AdS/CFT: From Shock Waves to Four-Point Functions,''
JHEP {\bf 0708}, 019 (2007).
[hep-th/0611122].
}

\lref\CornalbaZB{
  L.~Cornalba, M.~S.~Costa and J.~Penedones,
  ``Eikonal approximation in AdS/CFT: Resumming the gravitational loop expansion,''
JHEP {\bf 0709}, 037 (2007).
[arXiv:0707.0120 [hep-th]].
}

\lref\KabatPZ{
  D.~N.~Kabat,
  ``Validity of the Eikonal approximation,''
Comments Nucl.\ Part.\ Phys.\  {\bf 20}, 325 (1992).
[hep-th/9204103].
}

\lref\GiddingsGJ{
  S.~B.~Giddings and R.~A.~Porto,
  ``The Gravitational S-matrix,''
Phys.\ Rev.\ D {\bf 81}, 025002 (2010).
[arXiv:0908.0004 [hep-th]].
}

\lref\VenezianoER{
  G.~Veneziano,
  ``String-theoretic unitary S-matrix at the threshold of black-hole production,''
JHEP {\bf 0411}, 001 (2004).
[hep-th/0410166].
}

\lref\CostaMG{
  M.~S.~Costa, J.~Penedones, D.~Poland and S.~Rychkov,
  ``Spinning Conformal Correlators,''
JHEP {\bf 1111}, 071 (2011).
[arXiv:1107.3554 [hep-th]].
}

\lref\WeinbergKQ{
  S.~Weinberg and E.~Witten,
  ``Limits on Massless Particles,''
Phys.\ Lett.\ B {\bf 96}, 59 (1980).
}

\lref\CostaMG{
  M.~S.~Costa, J.~Penedones, D.~Poland and S.~Rychkov,
  ``Spinning Conformal Correlators,''
JHEP {\bf 1111}, 071 (2011).
[arXiv:1107.3554 [hep-th]].
}

\lref\BelitskyXXA{
  A.~V.~Belitsky, S.~Hohenegger, G.~P.~Korchemsky, E.~Sokatchev and A.~Zhiboedov,
  ``From correlation functions to event shapes,''
[arXiv:1309.0769 [hep-th]].
}

\lref\BuchelSK{
  A.~Buchel, J.~Escobedo, R.~C.~Myers, M.~F.~Paulos, A.~Sinha and M.~Smolkin,
  ``Holographic GB gravity in arbitrary dimensions,''
JHEP {\bf 1003}, 111 (2010).
[arXiv:0911.4257 [hep-th]].
}

\lref\ZhiboedovBM{
  A.~Zhiboedov,
  ``A note on three-point functions of conserved currents,''
[arXiv:1206.6370 [hep-th]].
}

\lref\OsbornCR{
  H.~Osborn and A.~C.~Petkou,
  ``Implications of conformal invariance in field theories for general dimensions,''
Annals Phys.\  {\bf 231}, 311 (1994).
[hep-th/9307010].
}

\lref\KomargodskiEK{
  Z.~Komargodski and A.~Zhiboedov,
  ``Convexity and Liberation at Large Spin,''
JHEP {\bf 1311}, 140 (2013).
[arXiv:1212.4103 [hep-th]].
}

\lref\MackJE{
  G.~Mack,
  ``All Unitary Ray Representations of the Conformal Group SU(2,2) with Positive Energy,''
Commun.\ Math.\ Phys.\  {\bf 55}, 1 (1977).
}

\lref\GrinsteinQK{
  B.~Grinstein, K.~A.~Intriligator and I.~Z.~Rothstein,
  ``Comments on Unparticles,''
Phys.\ Lett.\ B {\bf 662}, 367 (2008).
[arXiv:0801.1140 [hep-ph]].
}

\lref\KulaxiziJT{
  M.~Kulaxizi and A.~Parnachev,
  ``Energy Flux Positivity and Unitarity in CFTs,''
Phys.\ Rev.\ Lett.\  {\bf 106}, 011601 (2011).
[arXiv:1007.0553 [hep-th]].
}

\lref\HofmanAR{
  D.~M.~Hofman and J.~Maldacena,
  ``Conformal collider physics: Energy and charge correlations,''
JHEP {\bf 0805}, 012 (2008).
[arXiv:0803.1467 [hep-th]].
}

\lref\EpsteinBG{
  H.~Epstein, V.~Glaser and A.~Martin,
  ``Polynomial behaviour of scattering amplitudes at fixed momentum transfer in theories with local observables,''
Commun.\ Math.\ Phys.\  {\bf 13}, 257 (1969).
}

\lref\GiddingsGJ{
  S.~B.~Giddings and R.~A.~Porto,
  ``The Gravitational S-matrix,''
Phys.\ Rev.\ D {\bf 81}, 025002 (2010).
[arXiv:0908.0004 [hep-th]].
}

\lref\ZhiboedovOPA{
  A.~Zhiboedov,
  ``On Conformal Field Theories With Extremal a/c Values,''
JHEP {\bf 1404}, 038 (2014).
[arXiv:1304.6075 [hep-th]].
}

\lref\BelitskyXXA{
  A.~V.~Belitsky, S.~Hohenegger, G.~P.~Korchemsky, E.~Sokatchev and A.~Zhiboedov,
  ``From correlation functions to event shapes,''
Nucl.\ Phys.\ B {\bf 884}, 305 (2014).
[arXiv:1309.0769 [hep-th]].
}

\lref\RattazziPE{
  R.~Rattazzi, V.~S.~Rychkov, E.~Tonni and A.~Vichi,
  ``Bounding scalar operator dimensions in 4D CFT,''
JHEP {\bf 0812}, 031 (2008).
[arXiv:0807.0004 [hep-th]].
}

\lref\RychkovIJ{
  V.~S.~Rychkov and A.~Vichi,
  ``Universal Constraints on Conformal Operator Dimensions,''
Phys.\ Rev.\ D {\bf 80}, 045006 (2009).
[arXiv:0905.2211 [hep-th]].
}

\lref\WeinbergKQ{
  S.~Weinberg and E.~Witten,
  ``Limits on Massless Particles,''
Phys.\ Lett.\ B {\bf 96}, 59 (1980).
}

\lref\MinwallaKA{
  S.~Minwalla,
  ``Restrictions imposed by superconformal invariance on quantum field theories,''
Adv.\ Theor.\ Math.\ Phys.\  {\bf 2}, 781 (1998).
[hep-th/9712074].
}

\lref\ManoharTZ{
  A.~V.~Manohar,
  ``An Introduction to spin dependent deep inelastic scattering,''
In *Lake Louise 1992, Symmetry and spin in the standard model* 1-46.
[hep-ph/9204208].
}

\lref\HughesKF{
  V.~W.~Hughes and J.~Kuti,
  ``Internal Spin Structure of the Nucleon,''
Ann.\ Rev.\ Nucl.\ Part.\ Sci.\  {\bf 33}, 611 (1983).
}

\lref\BuchelSK{
  A.~Buchel, J.~Escobedo, R.~C.~Myers, M.~F.~Paulos, A.~Sinha and M.~Smolkin,
  ``Holographic GB gravity in arbitrary dimensions,''
JHEP {\bf 1003}, 111 (2010).
[arXiv:0911.4257 [hep-th]].
}

\lref\NachtmannMR{
  O.~Nachtmann,
  ``Positivity constraints for anomalous dimensions,''
Nucl.\ Phys.\ B {\bf 63}, 237 (1973).
}

\lref\ElkhidirWOA{
  E.~Elkhidir, D.~Karateev and M.~Serone,
  ``General Three-Point Functions in 4D CFT,''
JHEP {\bf 1501}, 133 (2015).
[arXiv:1412.1796 [hep-th]].
}

\lref\PolyakovGS{
  A.~M.~Polyakov,
  ``Nonhamiltonian approach to conformal quantum field theory,''
Zh.\ Eksp.\ Teor.\ Fiz.\  {\bf 66}, 23 (1974)..
}

\lref\NachtmannMR{
  O.~Nachtmann,
  ``Positivity constraints for anomalous dimensions,''
Nucl.\ Phys.\ B {\bf 63}, 237 (1973).
}

\lref\inpr{
 Work in Progress.
}

\lref\FitzpatrickYX{
  A.~L.~Fitzpatrick, J.~Kaplan, D.~Poland and D.~Simmons-Duffin,
  ``The Analytic Bootstrap and AdS Superhorizon Locality,''
JHEP {\bf 1312}, 004 (2013).
[arXiv:1212.3616 [hep-th]].
}

\lref\BuchelSK{
  A.~Buchel, J.~Escobedo, R.~C.~Myers, M.~F.~Paulos, A.~Sinha and M.~Smolkin,
  ``Holographic GB gravity in arbitrary dimensions,''
JHEP {\bf 1003}, 111 (2010).
[arXiv:0911.4257 [hep-th]].
}

\lref\HofmanUG{
  D.~M.~Hofman,
  ``Higher Derivative Gravity, Causality and Positivity of Energy in a UV complete QFT,''
Nucl.\ Phys.\ B {\bf 823}, 174 (2009).
[arXiv:0907.1625 [hep-th]].
}

\lref\HartmanLFA{
  T.~Hartman, S.~Jain and S.~Kundu,
  ``Causality Constraints in Conformal Field Theory,''
[arXiv:1509.00014 [hep-th]].
}

\lref\BassoZOA{
  B.~Basso, S.~Komatsu and P.~Vieira,
  ``Structure Constants and Integrable Bootstrap in Planar N=4 SYM Theory,''
[arXiv:1505.06745 [hep-th]].
}

\lref\HartmanLFA{
  T.~Hartman, S.~Jain and S.~Kundu,
  ``Causality Constraints in Conformal Field Theory,''
[arXiv:1509.00014 [hep-th]].
}

\lref\LiITL{
  D.~Li, D.~Meltzer and D.~Poland,
  ``Conformal Collider Physics from the Lightcone Bootstrap,''
[arXiv:1511.08025 [hep-th]].
}

\lref\RychkovET{
  S.~Rychkov,
  ``Conformal Bootstrap in Three Dimensions?,''
[arXiv:1111.2115 [hep-th]].
}

\lref\CamanhoAPA{
  X.~O.~Camanho, J.~D.~Edelstein, J.~Maldacena and A.~Zhiboedov,
  ``Causality Constraints on Corrections to the Graviton Three-Point Coupling,''
[arXiv:1407.5597 [hep-th]].
}

\lref\ZhiboedovOPA{
  A.~Zhiboedov,
  ``On Conformal Field Theories With Extremal a/c Values,''
JHEP {\bf 1404}, 038 (2014).
[arXiv:1304.6075 [hep-th]].
}

\lref\RattazziPE{
  R.~Rattazzi, V.~S.~Rychkov, E.~Tonni and A.~Vichi,
  ``Bounding scalar operator dimensions in 4D CFT,''
JHEP {\bf 0812}, 031 (2008).
[arXiv:0807.0004 [hep-th]].
}

\lref\MaldacenaJN{
  J.~Maldacena and A.~Zhiboedov,
  ``Constraining Conformal Field Theories with A Higher Spin Symmetry,''
J.\ Phys.\ A {\bf 46}, 214011 (2013).
[arXiv:1112.1016 [hep-th]].
}

\lref\MaldacenaNZ{
  J.~M.~Maldacena and G.~L.~Pimentel,
  ``On graviton non-Gaussianities during inflation,''
JHEP {\bf 1109}, 045 (2011).
[arXiv:1104.2846 [hep-th]].
}

\lref\GiombiRZ{
  S.~Giombi, S.~Prakash and X.~Yin,
  ``A Note on CFT Correlators in Three Dimensions,''
JHEP {\bf 1307}, 105 (2013).
[arXiv:1104.4317 [hep-th]].
}

\lref\MaldacenaIUA{
  J.~Maldacena, D.~Simmons-Duffin and A.~Zhiboedov,
  ``Looking for a bulk point,''
[arXiv:1509.03612 [hep-th]].
}

\lref\AdamsSV{
  A.~Adams, N.~Arkani-Hamed, S.~Dubovsky, A.~Nicolis and R.~Rattazzi,
  ``Causality, analyticity and an IR obstruction to UV completion,''
JHEP {\bf 0610}, 014 (2006).
[hep-th/0602178].
}

\lref\FordID{
  L.~H.~Ford,
  ``Constraints on negative energy fluxes,''
Phys.\ Rev.\ D {\bf 43}, 3972 (1991).
}

\lref\FordQV{
  L.~H.~Ford and T.~A.~Roman,
  ``The Quantum interest conjecture,''
Phys.\ Rev.\ D {\bf 60}, 104018 (1999).
[gr-qc/9901074].
}

\lref\BoussoWCA{
  R.~Bousso, Z.~Fisher, J.~Koeller, S.~Leichenauer and A.~C.~Wall,
  ``Proof of the Quantum Null Energy Condition,''
Phys.\ Rev.\ D {\bf 93}, no. 2, 024017 (2016).
[arXiv:1509.02542 [hep-th]].
}

\lref\BoussoMNA{
  R.~Bousso, Z.~Fisher, S.~Leichenauer and A.~C.~Wall,
  ``A Quantum Focussing Conjecture,''
[arXiv:1506.02669 [hep-th]].
}

\lref\FarnsworthHUM{
  K.~Farnsworth, M.~A.~Luty and V.~Prilepina,
  ``Positive Energy Conditions in 4D Conformal Field Theory,''
[arXiv:1512.01592 [hep-th]].
}

\lref\PolyakovYerevanNotes{
  A.~M.~Polyakov,
``Scale Invariance of Strong Interactions and its application to Lepton-Hadron Reactions,''
  Lecture Notes in {\it International School of High Energy Physics in Erevan, 23 November - 4 December 1971 (Chernogolovka 1972).}
  Acad. Sci. USSR, 1972
}

\lref\BzowskiQJA{
  A.~Bzowski and K.~Skenderis,
  ``Comments on scale and conformal invariance,''
JHEP {\bf 1408}, 027 (2014).
[arXiv:1402.3208 [hep-th]].
}

\lref\DymarskyZJA{
  A.~Dymarsky, K.~Farnsworth, Z.~Komargodski, M.~A.~Luty and V.~Prilepina,
  ``Scale Invariance, Conformality, and Generalized Free Fields,''
JHEP {\bf 1602}, 099 (2016).
[arXiv:1402.6322 [hep-th]].
}

\hfill TCDMATH-16-01
\Title{
\vbox{\baselineskip 6pt
}}
{\vbox{\centerline{Conformal Field Theories and Deep Inelastic Scattering}
}}\vskip.1in
 \centerline{
Zohar Komargodski,$^1$ Manuela Kulaxizi,$^2$ Andrei Parnachev,$^{2,3}$ Alexander Zhiboedov$^4$} \vskip.1in 
\centerline{\it $^1$
Weizmann Institute of Science, Rehovot 76100, Israel}
\centerline{\it $^2$
School of Mathematics, Trinity College Dublin, Dublin 2, Ireland}
\centerline{\it $^3$
Institute Lorentz, Leiden University, Leiden, The Netherlands}
\centerline{\it $^4$
Department of Physics, Harvard University, Cambridge, MA 02138, USA}

\vskip.9in \centerline{\bf Abstract} { We consider Deep Inelastic Scattering (DIS) thought experiments in unitary Conformal Field Theories (CFTs). 
We explore the implications of the standard dispersion relations for the OPE data. We derive positivity
constraints on the OPE coefficients of minimal-twist operators of even spin $s \geq 2$. In the case of $s=2$, when the leading-twist operator is the stress tensor, we reproduce the Hofman-Maldacena bounds. For $s>2$ the bounds are new.  
 }
 
\Date{January 2016}

\listtoc\writetoc
\vskip .5in \noindent

\newsec{Introduction and Summary}

Conformal field theories (CFTs) in $d$ spacetime dimensions are described first and foremost by correlation functions of local operators. The Operator Product Expansion (OPE) fixes these in terms of the spectrum of local operators and their three-point functions. Conformal symmetry determines the three-point functions up to a set of numbers.
The spectrum of unitary CFTs is constrained by unitarity bounds, which follow from the operator-state correspondence and the requirement that states have positive norm
 \refs{\MackJE\MinwallaKA-\GrinsteinQK}. 
There are, however, less obvious bounds coming from, for example,  positivity of energy correlators~\HofmanAR , deep inelastic scattering sum rules~\refs{\NachtmannMR , \KomargodskiEK},  and causality~\HartmanLFA. 

In the case of energy correlators \HofmanAR, one demands positivity of the energy flux at infinity integrated over all times.
 In the simplest case of a state created by a local operator with a given momentum this leads to  new constraints on the three-point functions of operators with spin and the stress-energy tensor of the type $\la {\cal O}^\dagger_{\mu_1 ... \mu_s} T_{\mu \nu} {\cal O}_{\nu_1 ... \nu_s}\ra$. 
The positivity of the integrated energy flux is a plausible assumption, but one may wonder whether there is
an independent argument for it. There have been a couple of proposals in the literature: in \HofmanUG\  the energy flux positivity has been derived from nontrivial assumptions about the OPE and the spectrum of non-local operators. Understanding the properties of these non-local operators and their OPE in unitary CFTs is an open problem. Another proposal has been put forward in~\KulaxiziJT, where the OPE of two stress-energy tensors has been extrapolated beyond the region of its validity to argue the energy flux positivity.

In
 \refs{\NachtmannMR , \KomargodskiEK}
 it was shown that by considering a setup where a particle with spin is scattered off
 a massive state, one can  relate (using the optical theorem) the positivity of the inclusive cross section (unitarity requires the cross section to be positive) with the OPE data, thereby placing constraints on the latter\foot{The assumption in~\KomargodskiEK\ involves the existence of a relevant operator which induces an RG flow
 terminating in a gapped phase; the scattering experiment involves the lightest particle in that gapped theory. Here we will argue that this additional structure is not necessary.}.
This  leads to the convexity property of the minimal twist operators which appear in the OPE of two Hermitian-conjugate operators.
In this paper we use a similar deep inelastic scattering (DIS) setup to derive the positivity of the energy flux and
related constraints on the OPE data for operators with spin. The idea of using DIS together with scale invariance is not new - for an example, see \PolyakovYerevanNotes.
We also discuss how to formulate the DIS experiment purely in a CFT without considering a flow to a gapped phase.

The results of our paper can be summarized as positivity constraints on the coefficients of the 
operator product expansion\foot{Three-point functions of operators with spin were analyzed in \CostaMG . $a_{s,m}^*$ are certain linear combinations of the structures from that paper as explained in detail in the main body of the present note. }
\eqn\ope{   {\cal O}_{j}^\dagger {\cal O}_j  \sim  \sum_m  a_{s,m}^*  {\cal O}_{\tau^*, s} +\ldots  \ ,
} 
where ${\cal O}_{\tau^*, s}$ is the minimal twist operator of even spin $s$ (the twist is defined as $\tau = \Delta -s$
where $\Delta$ is the conformal dimension and $s$ is the spin), the
index $m$ refers to the different tensor structures which appear in the DIS sum rules, and
the dots stand for the contribution of  higher twist operators. 
Then the coefficients  $a_{s,m}^*$ satisfy the following conditions:
\eqn\constrainta{\eqalign{
a_{2,m}^{*} \geq 0, &~~~m=0, ... ,j  \ , \cr
a_{s, m_1}^* a_{s,m_2}^* &\geq 0 \ , ~~~m_1, m_2 = 0, ... , j \ .
}}
In the case of $s=2$ the bounds above \constrainta\ are the familiar Hofman-Maldacena bounds \refs{\HofmanAR,\BuchelSK} because the minimal twist spin two operator is always a stress tensor.

In fact, we can obtain more general bounds by considering a four-point function of the type 
\eqn\fourpoint{
\la {\cal O}_{j}^\dagger {\cal O}_j \tilde {\cal O}_{\tilde j}^\dagger  \tilde {\cal O}_{\tilde j} \ra \ .
}
As before, we denote the minimal-twist operators which appear in the expansion of the correlator  \fourpoint\ in the s-channel
by  ${\cal O}_{\tau^*, s}$ and the corresponding OPE coefficients by  $a_{s,m}^*$ and $\tilde a_{s,m}^*$.
Then a more general set of constraints derived from DIS can be formulated as
\eqn\constraint{\eqalign{
a_{s, m_1}^* \tilde a_{s,m_2}^* &\geq 0 \ , ~~~m_1 = 0, ... , j,~~~ m_2= 0, ... , \tilde j \ .
}}

In the course of deriving the DIS sum rules, we {\it assumed} a certain behavior for the scattering amplitudes in the Regge limit. 
This translates into the lowest spin $s_c$ for which we can trust the sum rules. Thus, strictly speaking, our argument implies \constraint\ only for $s \geq s_c$, where $s_c$ is some unknown number which depends both on the theory and the external operators. However, when the external operators are energy momentum tensors, then there is evidence that $s_c=2$. We will discuss this point in detail below.

The ideas explained here have various applications for holographic theories, but we will pursue them elsewhere~\inpr.
One simple example that makes contact with~\HartmanLFA,\MaldacenaIUA\ is to consider a quartic scalar coupling in the bulk $\sim \lambda (\del\varphi)^4$. Denote the operator dual to $\varphi$ by $\CO$. The interaction $\sim \lambda (\del\varphi)^4$ shifts the dimension of the spin-two operator $\CO \del_\mu\del_\nu \CO $ to 
$ 2+2\Delta_\CO-\lambda$. Therefore, convexity is obeyed only if $\lambda>0$, as demanded by causality in the bulk \AdamsSV.

The rest of the paper is organized as follows. In section 2 we consider the DIS experiment with gravitons and derive constraints on the OPE coefficients of two stress tensors in a unitary CFT. In section 3 we generalize these considerations to the case of generic operators. In section 4 we derive a relation between the bounds obtained from the positivity of the energy flux and the DIS experiment. In section 5 we comment on how one can set up the DIS experiment without flowing to the gapped phase. Many technical details are collected in appendices.

\newsec{Deep Inelastic Scatering}

Deep Inelastic Scattering probes the internal structure of matter. The scattering process consists of bombarding a target with a highly energetic quantum and examining the final state. 
DIS was first used to probe the structure of hadronic particles. 
The setup is depicted in Fig.1. A lepton emits a virtual photon which strikes a hadron. In principle, to investigate the structure of the target $|P\rangle$ one may shoot different particles at it. A natural choice are particles which couple to conserved currents. The options depend on the theory and the symmetries it preserves. A universal choice to consider is the graviton. We can couple the stress-energy tensor of the theory to the background graviton and perform the DIS experiment.  More generally, we can couple a source to any operator of the theory.
\midinsert\bigskip{\vbox{{\epsfxsize=2in
        \nobreak
    \centerline{\epsfbox{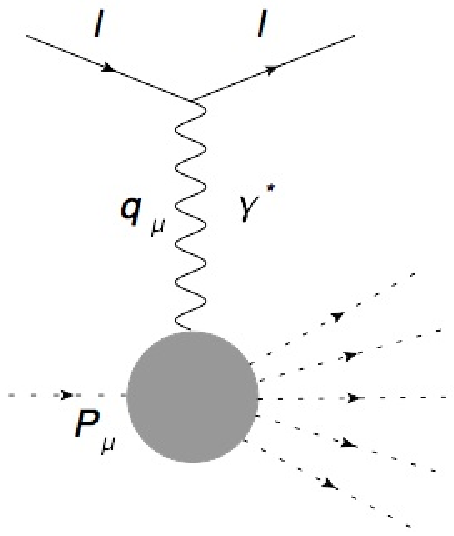}}
        \nobreak\bigskip
    {\raggedright\it \vbox{
{\bf Fig 1.}
{\it  A lepton emits a virtual photon which strikes a hadron. The hadron breaks up into a complicated final state.
}}}}}}
\bigskip\endinsert
\noindent

We also have to specify the state $|P\rangle $. For that we imagine that our theory is gapped and we denote with $|P\rangle $ the lightest, massive, one-particle state in the system which we assume to be a scalar.

In the standard treatment of DIS one can relate the deep Euclidean (i.e. ultraviolet) data to the positive-definite total cross section using dispersion relations. While our presentation is aimed to be self-contained, one can consult, for instance, the reviews~\refs{\HughesKF,\ManoharTZ}.

It was already demonstrated in~\refs{\NachtmannMR,\KomargodskiEK} that the ideas of DIS can lead to nontrivial consequences for unitary CFTs. 
There it was argued that the minimal twist of operators which appear in the OPE of Hermitian conjugate operators is a monotonic, non-concave function of spin starting from some $s \geq s_c$.

In what follows we will discuss the DIS experiment with gravitons and restrict to the case of a scalar target $\ket{P}$. Later we will argue that it is not necessary to make this series of assumptions. In the meanwhile, we make these assumptions in order to simplify the presentation.

\subsec{A DIS Experiment with Gravitons}

Let us consider the DIS experiment for the case of the stress-energy tensor operator $T_{\mu\nu}(x)$. A background graviton $\delta g_{\mu\nu}(x)$ couples to the theory via $\sim\int d^dx \ T^{\mu\nu}(x) \delta g_{\mu\nu}(x)$. We imagine that some physical particle emits an off-shell graviton which strikes a state of the theory. So we have in mind the setup of Fig.1, only with the photon replaced by the graviton.

A useful intermediate object to consider is the ``DIS amplitude.'' For that we imagine an exclusive process, where the graviton strikes the state $|P\rangle $ and the out states are again a graviton (with the same polarization and momentum) and the same initial state, $\langle P| $. This is depicted in Fig.2. 
\midinsert\bigskip{\vbox{{\epsfxsize=2in
        \nobreak
    \centerline{\epsfbox{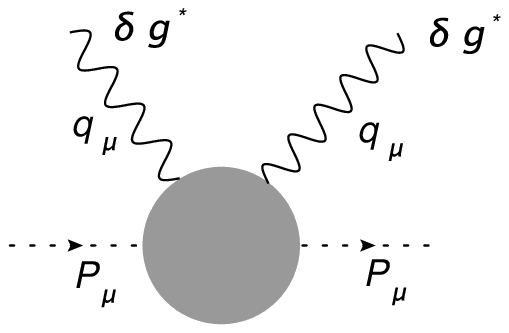}}
        \nobreak\bigskip
    {\raggedright\it \vbox{
{\bf Fig 2.}
{\it  The Deep Inelastic Scattering Amplitude. $\delta g^*$ stands for a virtual graviton with momentum $q_\mu$.
}}}}}}
\bigskip\endinsert
\noindent

The amplitude for the ``graviton-DIS'' process depicted in Fig.2 is given by
\eqn\AdefT{
\AA(q_\mu,P_\nu)=\int d^4y e^{-i q y} \vev{P|\TT \left(T(\epsilon^\star,y) T(\epsilon,0)\right)|P} \ , 
}
where the momentum of the target $|P\rangle$ is denoted by $P_\mu$; $T(\epsilon,y) \equiv T_{\mu\nu}(y)\epsilon^{\mu\nu}$ and $\epsilon^{\mu\nu}$ is a polarization 
tensor ($\epsilon^\star$ is the conjugate polarization tensor). We can shift $\epsilon^{\mu\nu}\rightarrow \epsilon^{\mu\nu}+q^\nu l^\mu+q^\mu l^\nu$ with arbitrary $l^\nu$. This would not affect the two-point function in the vacuum because of energy-momentum conservation. But here we are dealing with a two-point function in a nontrivial state so contact terms may contribute. 
We therefore do not impose $\eps.q=0$.  However, note that if we were to take $\epsilon^{\mu\nu}\sim \eta^{\mu\nu}$ then we would be studying the scattering of the conformal mode of the metric, i.e. the dilaton. These scattering amplitudes are suppressed at large $q$ because the trace of the energy momentum tensor vanishes in a conformal field theory. We therefore take the tensor $\epsilon^{\mu\nu}$ to be traceless.

We imagine a general massive (non-conformal, gapped) theory, in which the lightest state is $|P\rangle $. The above amplitude depends on the mass scales of the theory, on the polarisation $\epsilon^{\mu\nu}$, and on two kinematical invariants, i.e., $q^2$ and $x={q^2\over 2 q .P}$. We promote $x$ to a complex variable and study the amplitude for fixed {\it spacelike} momentum $q^2>0$. Since $|P\rangle$ is the lightest particle of the theory (which we assume to be a scalar for simplicity), the above amplitude will have a branch cut discontinuity for $-1\leq x \leq 1$, as depicted in Fig.3. The optical theorem relates the discontinuity across the cut in the $x$-plane to the square of the forward amplitude, which is positive-definite. 
\midinsert\bigskip{\vbox{{\epsfxsize=3in
        \nobreak
    \centerline{\epsfbox{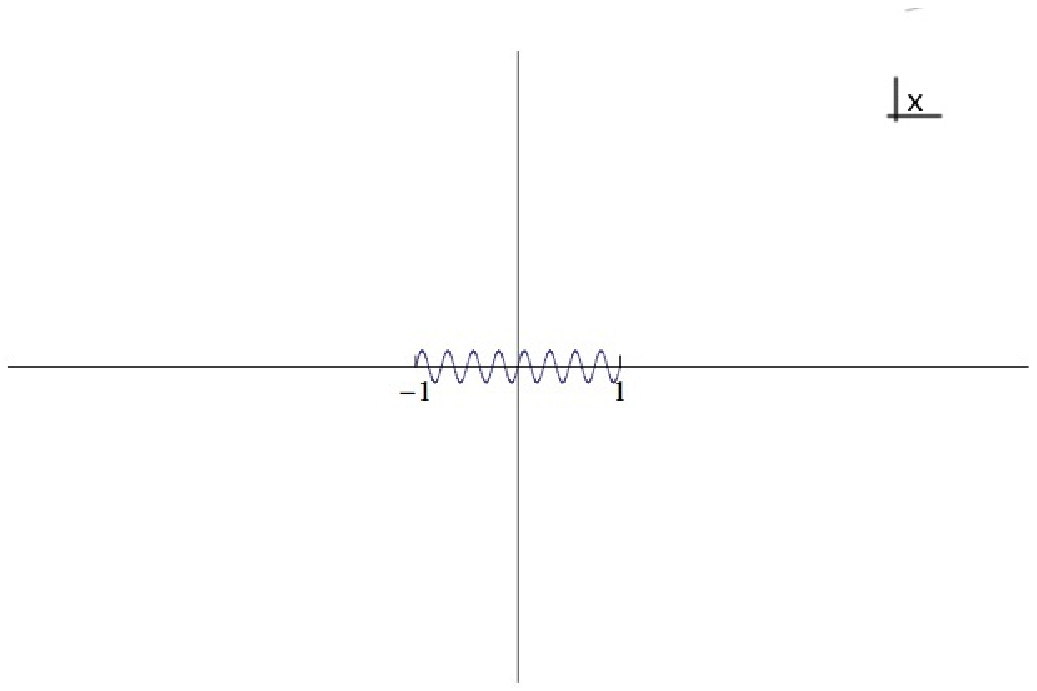}}
        \nobreak\bigskip
    {\raggedright\it \vbox{
{\bf Fig. 3.}
{\it  The analytic structure in the $x$-plane.
}}}}}}
\bigskip\endinsert
\noindent

For large (compared to the mass scale) and space-like $q^2>0$ we can compute the DIS amplitude \AdefT\ with the help of the OPE, which is determined in the ultraviolet Conformal Field Theory. The resulting expression is a series expansion around $x\rightarrow \infty$, valid for fixed and large $q^2>0$. To isolate the coefficient of the $s$-th power of $x$ in the expansion, one computes the ``$s$-moment" defined as $\mu_s(q^2)=\oint dx x^{s-1} \AA(x,q^2)$. As long as the amplitude vanishes sufficiently fast for small $x$, we can pull the contour from infinity to the branch cut and write
\eqn\ct{\oint dx \ x^{s-1} \AA(x,q^2)=2\int_0^1 dx \ x^{s-1} {\rm Im}[\AA(x,q^2) ]\, ,}
where we assumed that $s$ is even. For odd $s$ the contribution from the left and the right cut cancel each other.

For~\ct\ to be valid for all $s\geq 2$ we need to assume that 
\eqn\Asmallxi{\lim_{x\rightarrow 0} \AA(q^2,x) < x^{-2}\,~.}
In general we only know that $\AA$ is bounded by some $x^{-N}$ in this limit. (This is discussed in~\EpsteinBG; for a recent discussion and references see also \GiddingsGJ.) However, there are some pieces of evidence that~\Asmallxi\ indeed holds for graviton deep inelastic scattering. One is that the convexity theorems derived from it in~\KomargodskiEK\ hold in all known examples. The other piece of evidence is that, as we will show below, by assuming~\Asmallxi\ we get precisely the bounds of~\HofmanAR\ if we focus on $s=2$. We will therefore take~\Asmallxi\ as an assumption in this section and revisit it in the next section when we discuss more general DIS gedanken experiments.\foot{As mentioned in the introduction, the bounds following from $s=2$ constrain the allowed effective theories in AdS. For example, the bound on the sign of $\lambda$ in the $\lambda (\pa\phi)^4$ theory in AdS recently discussed in \refs{\HartmanLFA,\MaldacenaIUA} immediately follows from the convexity of anomalous dimensions, assuming the $s=2$ sum rule converges.}

Unitarity implies
\eqn\imA{{\rm Im}[\AA(x,q^2)]\geq 0\,,}
which leads, via \ct, to
\eqn\uniA{\oint dx x^{s-1} \AA(x,q^2)\geq 0\,,}
imposing positivity relations on the coefficients of the OPE. These constraints are in addition to the non-concavity of the minimal twist function.

\subsec{OPE in DIS Kinematics}

Our objective is to evaluate \AdefT\ with the help of the OPE and investigate the positivity constraints one obtains from \uniA. We start with the operator product expansion for two energy-momentum tensors
\eqn\TTope{T(\epsilon^\star, y) T(\epsilon,0)=\sum_{s=0,2,4\cdots}\sum_\alpha \hat{f}_s^{(\alpha),{\mu_1\cdots \mu_s}}(y,\epsilon,\epsilon^\star)\OO^{(\alpha)}_{\mu_1\cdots\mu_s}(0)}
where $s$ denotes the spin of the operator and $\alpha$ labels operators of the same spin. Actually there could be operators in other representations in~\TTope, for example, operators in mixed symmetric-antisymmetric representations (see e.g.~\ElkhidirWOA). Since we are ultimately interested in using the OPE to evaluate \AdefT, we can ignore the representations which have some of their indices anti-symmetrized because the corresponding expectation values in the (scalar) state $| P \ra$ vanish. For a similar reason,  we do not include descendants in~\TTope; they give a vanishing contribution because $\pa_{\mu}\la P| {\cal O}(x)| P \ra = 0$.

For the operators $\OO_{\mu_1\mu_2\cdots\mu_s}^{(\alpha)}$ in the even $s$ symmetric traceless representation which appear in the OPE~\TTope,  the expectation values of $\OO_{\mu_1\mu_2\cdots\mu_s}^{(\alpha)}$ in the state $|P\rangle$ are parametrized as
\eqn\expO{\langle P|\OO^{(\alpha)}_{\mu_1\mu_2\cdots\mu_s}(0)|P\rangle=B_s^{(\alpha)} P_{\mu_1}P_{\mu_2}\cdots P_{\mu_s}-\cdots\,, }
where the $B_s^{(\alpha)}$ are some dimensionful coefficients and the dots stand for trace terms (terms involving the metric tensor), which we will not need to specify. For example, in the case of the stress tensor expectation value in a one-particle state, we famously have~\WeinbergKQ\foot{We normalize the one-particle states as follows $\la P' | P \ra = (2 \pi)^{d-1} E_{\vec P} \delta^{(d-1)}(\vec P' - \vec P)$.} \eqn\emvev{\langle P| T_{\mu_1\mu_2}(0)|P\rangle = P_{\mu_1} P_{\mu_2}~.} Therefore, the corresponding coefficient $B_{T}$ is determined to be 1. 

Conformal symmetry fixes the form of the leading OPE coefficients for small enough $y$ to be\foot{Here $\epsilon^\star .\epsilon = \epsilon^\star_{\alpha \beta}\epsilon^{\alpha \beta}$ and $(\epsilon^\star .\epsilon)^{\lambda\kappa} = \epsilon^{\star \lambda}_{\  \  \alpha} \epsilon^{\kappa \alpha}$.}
\eqn\fdef{\eqalign{  & \,\,  \hat{f}^{(\alpha)}_{\mu_1\cdots\mu_s}(\epsilon^\ast,\epsilon,y)=y^{\left(\tau_s^{(\alpha)} -2 d\right)}\biggl[ 
 \hat{a}_{s,0}^{(\alpha)}\, (\epsilon^\star .\epsilon) y_{\mu_1}\cdots y_{\mu_s}+\hat{a}_{s,1}^{(\alpha)} \, (\epsilon^\star .\epsilon)^{\lambda\kappa}y_\lambda y_{\kappa} y_{\mu_1}\cdots y_{\mu_s} (y^2)^{-1}+\cr 
&+\hat{a}_{s,2}^{(\alpha)} \,((\epsilon^\star)^{\kappa_1\kappa_2} y_{\kappa_1}y_{\kappa_2}) (\epsilon^{\lambda_1\lambda_2} y_{\lambda_1}y_{\lambda_2}) y_{\mu_1}\cdots y_{\mu_s} (y^2)^{-2} + ...\biggr]
\, ,}}
where the dots denote  terms which contain polarization tensors with non-contracted indices
as well as terms subleading in powers of $y$. Both of these will turn out to be sub-leading in the kinematics we are considering.

We now substitute \fdef\ in \AdefT\ and take the Fourier transform, leading to
\eqn\fourierope{\eqalign{&\AA(q_\mu, P_\mu)=\sum_{s=0,2,4,\cdots}\sum_{\alpha}   (\epsilon^\star .\epsilon) \left(\hat{a}_{s,0}^{(\alpha)} B_s^{(\alpha)}\left(i \,{\p\over \p q} .\,P\right)^s-\rm{traces}\right) f_{s,0}^{(\alpha)}(q)+\cr
&+(\epsilon^\star .\epsilon)^{\lambda\kappa}\left(i {\p\over \p q^\lambda}\right)\left(i {\p\over \p q^\kappa}\right)  \left(\hat{a}_{s,1}^{(\alpha)} B_s^{(\alpha)}\left(i\,{\p\over \p q}.\,P\right)^{s}-\rm{traces}\right) f_{s,1}^{(\alpha)}(q)+\cr
&+\left( \,{\p\over \p q^{\lambda_1}}{\p\over \p q^{\lambda_2}}\,(\epsilon^\star)^{\lambda_1\lambda_2}\right)\left( \,{\p\over \p q^{\kappa_1}}{\p\over \p q^{\kappa_2}}\epsilon^{\kappa_1\kappa_2}\right)  \left(\hat{a}_{s,2}^{(\alpha)} B_s^{(\alpha)}\left(i\,{\p\over \p q}.\,P\right)^{s}-\rm{traces}\right) f_{s,2}^{(\alpha)}(q)+\cdots }\,,}
Here the functions $f^{(\alpha)}_{s,m}(q)$ are Fourier transfomations of the ``Feynman" propagators, defined as follows
\eqn\fhatdef{\eqalign{f^{(\alpha)}_{s,m}(q)&=\int d^d y \, e^{-i q y}\,  (y^2+i\varepsilon)^{  {1\over 2}\tau_s^{(\alpha)} -d-m     } 
\,,}}
and ``traces" stands for terms of the form
$P^{2 n}\left(P .{\p\over \p q}\right)^{s-2 n} \left({\p\over\p q}.{\p\over \p q}\right)^n$ and $2\leq 2 n\leq  s$. 
We will soon see that these terms are negligible in the limit we consider.

At this point it is convenient to express the amplitude in terms of the kinematical invariants, $q^2,\, x\equiv {q^2\over 2 P.q}$. We are interested in the regime of large spacelike $q^2>0$ but we work to all orders in $x$. Therefore, for a given power of $x$ we keep only the leading terms in the limit $q^2\rightarrow \infty$. We obtain:
\eqn\AT{\eqalign{\AA(q^2,\, x)&=\sum_{s} (q^2)^{-\tau^{\ast}_{s,0}/2+d/2}
C_{s,0}^{\ast} x^{-s} \left(\epsilon^\star .\epsilon\right)^2\cr
&+\sum_{s} (q^2)^{-\tau^{\ast}_{s,1}/2+d/2} C_{s,1}^{\ast} x^{-s}{(\epsilon^\star .\epsilon)^{\lambda\kappa}q_\lambda q_\kappa\over q^2}\cr
&+\sum_{s} (q^2)^{-\tau^{\ast}_{s,2}/2+d/2} C_{s,2}^{\ast} x^{-s} {(\epsilon^\star)^{\lambda_1\lambda_2} q_{\lambda_1}q_{\lambda_2}\epsilon^{\kappa_1\kappa_2} q_{\kappa_1}q_{\kappa_2}\over (q^2)^2 }\, ,
}}
where $\tau_{s,i}^\ast$ denotes the twist of the minimal twist operator which contributes to the corresponding polarization tensor structure.
A priori we do not have  to impose $\tau_{s,i}^\ast = \tau_s^\ast$, but generically we do expect this to be the case since there is no symmetry principle that sets some of the tensor structures to zero. Below we assume that 
\eqn\twistnongen{
\tau_{s,m}^\ast = \tau_s^\ast \ ,
} unless stated otherwise\foot{One may worry that the large momentum limit of the DIS amplitude is not correctly captured by the Fourier transform of the OPE \refs{\DymarskyZJA,\BzowskiQJA}. We expect this issue not to be relevant here, because the terms which dominate over the Fourier transform of the OPE in the large momentum limit come from ``semi-local terms" \BzowskiQJA\ in position space. It would be interesting to show that this is indeed the case.}.

The ``trace" terms have been consistently neglected by invoking the monotonicity of the twists \KomargodskiEK. Similarly, one can verify that terms containing $\eps.P$ are irrelevant for our consideration. Among the set of operators of a given spin $s$, only the one with the smallest twist, $\tau_s^*$, has been retained in \AT. The corresponding coefficients, $C_{s,m}^*$, are given by 
\eqn\nfStress{\eqalign{C_{s,m}^*&= 2^{\tau^*_s - d - 2 m} \pi^{{d \over 2}} {\Gamma({\tau^*_s - d\over 2} + m + s) \over \Gamma(m + {2 d - \tau^*_s\over 2} ) } B_s^* a^*_{s,m},
}}
which can be derived using the Fourier transform~\fhatdef. Explicit expressions for the $a^*_{s,m}$ in terms of the $\hat{a}^*_{s,m}$ which appear in \fourierope\ are given in Appendix A.

As long as we can deform the contour in the complex plane as explained above \ct, we can substitute \AT\ into \uniA\ to obtain various positivity relations as required by unitarity, i.e.
\eqn\uniC{C_{s,m}^\ast\geq 0, \qquad m=0,1,2\,.}
These three inequalities for each spin are achieved by judicious choices of the polarization tensor. First, we choose a convenient reference frame for the space-like momentum $q^\mu=(0,0,\cdots,0,k)$. 
We then organize the polarization tensor $\epsilon_{\mu\nu}$ according to its properties under the subgroup of rotations which leave $q^\mu$ invariant. There are three possibilities and each produces a single constraint 
\item{$\bullet$} We can take $\epsilon_{01}=\epsilon_{10}=1$ and let all the other components vanish.  Then only the first line in~\AT\ remains. 
\item{$\bullet$} We take $\epsilon_{0 1}=\epsilon_{ 10}=\epsilon_{1(d-1)}=\epsilon_{(d-1)1}=1$ and all the other components vanish. 
Only the second line in~\AT\ remains non-zero. 
\item{$\bullet$} We take $\epsilon_{00}=\epsilon_{(d-1)(d-1)}=\epsilon_{0(d-1)}=\epsilon_{(d-1)0}=1$ with the rest of the components set to zero. In this case,  only the last line in~\AT\ is non-vanishing.

\bigskip

It is instructive to consider in detail the case $s=2$. In this case, the operator of the smallest twist is none other but the stress energy tensor. Unitarity sets a lower bound on the twist of all spin $s$ operators $\tau_s\geq d-2$ (and when the inequality is saturated we get a conserved current)\refs{\MackJE,\GrinsteinQK}. Hence, the energy-momentum tensor is the minimal twist operator with $s=2$ unless the theory has more than one conserved spin 2 current. For the energy-momentum tensor, we know from~\emvev\ that $B_{T} = 1$. It follows that~\uniC\ directly imposes bounds on the OPE coefficients of the CFT.

Remarkably, these bounds coincide with the energy flux constraints obtained in~\HofmanAR. To make this explicit, we should relate the $a_{T,m}$ to the independent OPE coefficients of $TT\sim T$ using the formalism of \refs{\CostaMG,\OsbornCR}. A similar computation in $d=4$ was done in \KulaxiziJT. For generic $d$ we get\foot{See appendix B for details on the derivation of these constraints.}
\eqn\CT{\eqalign{a_{T,0}&= -{d (2 b-c)+ a(d^2+4 d-4)\over 4(-2 b-c(1+d)+a(-6 +d+d^2))} \sim n_{v} \geq 0 \ , \cr
a_{T,1}&=\,\,{1\over 8} \,\, {a(d^2+6 d-8)-b(2-3 d)-2 d c\over (-2 b -c(1+d)+a(d^2+d-6))} \sim n_{f} \geq 0 \ , \cr
a_{T,2}&= \,- \, {1\over 32}\,\,  {d(d-2)(4a+2 b-c)\over (a(d^2+d-6)-2 b-c(1+d))} \sim n_{s} \geq 0 ,}}
where $(a,b,c)$ denote the parameters which determine the three-point function of the stress-energy tensor in the notations of \OsbornCR, and $(n_{s}, n_{f} , n_{v})$ in the basis of structures generated by free field theories \BuchelSK . \CT\ holds in any $d\geq 4$. In $d=4$, it yields the familiar expressions in four dimensions \HofmanAR. 

In $d=3$ dimensions, there are only two independent conformal free theories (those of free scalars and of free fermions) and the number of independent parameters in the three-point function of the stress energy tensor is accordingly reduced to two. In this case, explicit computation leads to two constraints $n_s\geq 0,\,\, n_f\geq 0$.\foot{Two out of the three structures in \AT\  yield $n_s\geq 0$ and the other one, $n_f\geq 0$. In three dimensions, an additional parity odd structure in the three-point function of the stress energy tensor is allowed \refs{\MaldacenaNZ,\GiombiRZ}. Here and in the rest of this paper, we restricted the discussion to parity even structures.}

Deep Inelastic Scattering allows for a clean separation between infrared physics and ultraviolet physics. This is a key ingredient in our arguments. In~\KulaxiziJT\ an attempt to use the OPE beyond its regime of validity has been discussed. We are circumventing this conceptual difficulty by the DIS analysis, which relates ultraviolet and infrared data by a contour argument.

Let us now discuss the case of higher spins, $s>2$. 
As explained in \refs{\KomargodskiEK,\FitzpatrickYX,\LiITL} in this case $d-2 \leq \tau^*_s < 2 (d-2)$ and because of this, the ratio of gamma functions that appears in $C_{s,m}^*$ is positive-definite. 
For spins $s>2$ we do not know the sign of $B_s^*$ but we can still get some mileage out the constraints above since $B_{s}^*$ does not depend on $m$. Assuming that for the minimal twist operator $B_s^* \neq 0$, we get an infinite set of new bounds for unitary CFTs
\eqn\newbounds{
a_{s,m_1}^*  a_{s,m_2}^*  \geq 0 .
}

This product appears naturally in the OPE of the four-point function of stress energy tensors. For this reason it seems reasonable to hope that the prediction~\newbounds\ can be tested in future studies of the conformal bootstrap for operators with spin.

\newsec{Deep Inelastic Scattering for Generic Operators}

In the previous section we considered the DIS of gravitons which couple to the stress tensor $T_{\mu \nu}$. One can naturally generalize this to any source that couples to some operator 
$\OO_j(\epsilon,\, x)$,\foot{Here $\OO_j(\epsilon,\, x) = \OO_{\mu_1 .. . \mu_j}(x) \eps^{\mu_1 ... \mu_j}$. } which is a symmetric, traceless CFT operator of spin $j$ and conformal dimension $\Delta_\OO$ that satisfies the unitarity bound $\Delta_\OO-j\geq d-2$.

For a generic external operator $\OO_j(\epsilon,\, x)$ we do not commit on the rate of the decay of the amplitude for small $x$ (except that it is bounded by some power). We will be more precise about this issue below.

The DIS amplitude of interest is 
\eqn\Adefgenj{\AA(q,\,P)=\int d^dy e^{-i q y} \vev{P | \TT \left(\OO^{\dagger}_j(\epsilon^\star, y)\OO_j(\epsilon,0)\right)|P }~. }
It is convenient to choose the polarization tensor as follows \CostaMG
\eqn\polariz{
\epsilon^{a_1...a_j}=\epsilon^{a_1}\cdots \epsilon^{a_j}~ , 
}
where $\epsilon^2=0$.  So we consider
\eqn\OOjdef{
\OO_j(\epsilon, y)\equiv \OO_{a_1\cdots a_j}(y) \epsilon^{a_1}\cdots \epsilon^{a_j}~.
}
By a straightforward generalization of the previous analysis, we obtain an expression for the  DIS amplitude \Adefgenj\ in the limit of large $q^2$,
\eqn\Aj{\AA(q^2,x)=\sum_s (q^2)^{-\tau^\ast/2+\Delta_\OO -d/2} x^{-s} \sum_{m=0}^{j}C_{s,m}^\ast \, (\epsilon^\ast .\epsilon)^{j-m} {(\epsilon^\ast .q)^m (\epsilon .q)^m\over (q^2)^m} \, ,}
where the constants are defined as $C_{s,m}^{\ast}\propto a_{s,m}^\ast B_s^\ast$ with a proportionality coefficient derived from the Fourier transform (we will give explicit expressions soon) and the asterisk represents the lowest twist for each spin-$s$ operator in the OPE. 
Substituting \Aj\ into \uniA\ leads (after appropriate choices of the polarisation tensor) to positivity constraints on the coefficients of the expansion
\eqn\ctj{C_{s,m}^\ast \geq 0,\qquad \qquad m=0,1,\cdots, j~.}
Focusing upon the stress energy operator ($s=2$) on the right hand side of~\Aj, we find positivity requirements for the OPE coefficients in unitary CFTs. There are in total $(j+1)$-positivity conditions.

It is time to discuss to what extent we can trust~\ctj\ for all $s\geq 2$. The validity of \uniA\ is  dependent upon the  behaviour of the DIS amplitude for fixed $q^2$ and small $x$, or equivalently, large $\nu=2 P.q$. If we assume 
\eqn\Asmallx{\lim_{x\rightarrow 0} \AA(q^2,x) \leq x^{-N}}
for some integer $N$,  the DIS sum rules and the bounds~\ctj\ would be justified for $s\geq N$.

We can try to obtain some information about $N$ indirectly as follows. 
As previously mentioned, $C_{s,m}^\ast$ is proportional to the OPE coefficient times the expectation value $B_{s}^\ast$ up to an overall number derived from  a Fourier transform, as in \fhatdef. For the case at hand, of generic external operators of spin-$j$ and conformal dimension $\Delta_\OO$, the relevant Fourier transform is
\eqn\ftg{\int d^d y \, e^{i q y} (y^2+i\varepsilon)^{{1\over 2}\tau^{\ast}_s -\Delta_\OO-m} ={\pi^{{d\over 2}}\Gamma[d/2+\tau^\ast/2-\Delta_\OO-m] \over \Gamma[-\tau^\ast/2+\Delta_\OO+m]}\left(q^2/4-i\epsilon\right)^{-\tau_\ast/2+\Delta_\OO+m-d/2}\,.}
The precise expression for $C_{s,m}$ in terms of $B_s^\ast$ and the OPE coefficients $a_{s,m}^\ast$ is
\eqn\nf{\eqalign{
C_{s,m}&= 4^{-\beta} {\pi^{{d\over 2}} \Gamma[s+ m + {\tau^\ast\over 2} + {d \over 2} -\Delta_{\OO}]\over \Gamma[\Delta_\OO+m -{\tau^\ast\over 2}]} \,\, a^\ast_{s,m}\, B_s^\ast\,,\cr
\beta&=-{\tau^\ast\over 2}+\Delta_\OO+m-d/2
}}
and is obtained by differentiating \ftg\ $(s+2 m)$--times with respect to $q_\mu$. The $a_{s,m}^*$ are specific linear combinations of the position space $\hat{a}_{s,m}^*$, similarly to what happens in the graviton DIS.

Let us consider now the case of the stress tensor exchange, i.e., $\tau^\ast_T=d-2,\, s=2$. In this case $B_T = 1$ which leads to $a_{T,m}\geq 0$ as long as the numerical factor in \nf\ is positive definite. For $\tau^\ast_T=d-2$ and $s=2$  the arguments of the gamma functions in the numerator/denominator of \nf\ are equal to $d+m+1-\Delta_\OO$ and $\Delta_\OO+m+1-d/2$ respectively. 
The latter is positive definite by unitarity but the former is not necessarily positive. We get that it is positive-definite only for\foot{When $(d-\Delta_\OO+1)$ is a negative integer the Fourier transform of the integral should be regulated by adding a local term to cancel the $\Gamma$-function pole. The result for the overal coefficient is still a number of alternating sign.}
\eqn\deltabound{
\Delta_\OO \leq d+1 .
}

Equivalently, for $\Delta_\OO > d+ 1$ the Fourier-transform above is divergent and we define it by an analytic continuation. Assuming that the energy flux bounds still hold, we would get an apparent contradiction for the $\Delta_\OO$ for which the ratio of the $\Gamma$-functions changes sign (see appendix C for an explicit computation in the case of scalar, external operators). We think that this signals the need of subtraction in the sum rule.\foot{It can be easily seen that doing subtractions in the $x$ sum rule used in the previous section, automatically projects out all low spin operators from the OPE.} We consider the example of a free scalar field in appendix D, where the scenario just described is explicitly realized. 
Summarizing, for generic operators of conformal dimension $\Delta_\OO$ and spin-$j$, we expect to trust the $x$-sum rule and the derived constraints \ctj, in the spin-$s$ sector with $s\geq s_c\geq  \Delta_\OO-{\tau^*_{s_c}\over 2}-{d\over 2}$, where $s_c$ is the first spin for which it holds that $s\geq  \Delta_\OO-{\tau^*_s\over 2}-{d\over 2}$.

\newsec{DIS vs Energy Correlator: Are The Constraints Always Equivalent?}

In this section we show that the constraints one gets from the positivity of the energy flux in a state produced by a given local operator ${\cal O}_j$ of spin $j$  imply the constraints obtained from the DIS $s=2$ sum rule. More precisely, for the case of external operators which are conserved currents, we show that the constraints derived from the DIS sum rule and those obtained from the positivity of the energy correlators, are equivalent. On the other hand, for generic operators, the energy correlators constraints are stronger than the ones which follow from the DIS $s=2$ sum rule. The bounds that are associated to $s>2$ in DIS 
do not follow in any simple way from the positivity of the energy flux.

Consider now the energy flux operator, defined as in \refs{\HofmanAR},
\eqn\efdef{\EE(n)=\lim_{r \rightarrow \infty} r^{d-2} \int_{-\infty}^{\infty} d t \ T_{0 i} n^i (t, r n^i) \, ,
}
and $n = (1, \vec n)$ or equivalently one can define the calorimeter operator in a manifestly covariant way \ZhiboedovOPA.
The expectation value of the energy flux on the state 
\eqn\efstate{\ket{\OO_j(\epsilon, k)} =\int d^d y\, e^{i k x} \OO_j(\epsilon, x) \ket{0}\,,}
obtained by acting with the operator $\OO_j$ carrying momentum $k$ on the vacuum, is fixed
by rotational invariance up to a few parameters;
\eqn\efgenj{
\vev{\EE(n)}_{{\cal O}.\eps (k)} \sim   \vev{\OO_j(\epsilon , k)|\EE(n)|\OO_j(\epsilon , k)} ={(k^2)^{\Delta} \over  (k.n)^{d-1}} \,\sum_{\ell=0}^{j} D_\ell (\epsilon^\ast.\epsilon)^{j - \ell}{(\epsilon^\ast . k)^\ell (\epsilon. k)^\ell \over \left( k^{2} \right)^{\ell} }\,.
}
Here we imposed the transversality condition $\epsilon.n=0$. Notice that usually the polarization tensor is chosen such that $\eps.k=0$ (see for example, \HofmanAR), however for the purpose of  comparison with DIS, the choice above is more convenient. Conformal invariance determines the three-point correlation functions up to a few numbers and thus, the $D_{\ell}$ can be expressed as linear combinations of those numbers. Requiring positivity of the energy, $\vev{\EE(n)}\geq 0$, leads to $(j+1)$ linear constraints on the parameters $D_\ell \geq 0$ or, equivalently, on the constants which determine the three-point functions $\vev{\OO T\OO}$.

Below we show that the constraints obtained from the computation above in the $\eps.n=0$  ``gauge'' are identical to the ones derived from DIS, assuming that we can trust the $s=2$ dispersion relation integral.

\subsec{Computing the Energy Correlator}

Consider the three-point function $\la {\cal O}^{\dagger} (x_1, \eps^*) T_{\mu \nu}(x_2)  {\cal O}(x_3, \eps)\ra$. 
Together with the two-point function $\la {\cal O}^\dagger(x,\eps^*) {\cal O}(x,\eps)\ra$, it can be used to compute two objects: the one-point energy correlator and the OPE coefficient in
${\cal O}^{\dagger} (x, \eps^*) {\cal O}(0, \eps) \sim C^{\mu \nu} (x, \eps) T_{\mu \nu}(0)$. The letter is useful to obtain the DIS constraints as discussed in the previous sections. 
It was observed in section 2 that in some cases the constraints obtained via the two methods coincide.

In this section we show that the two always produce the same constraints provided that $\eps.n=0$. First, we consider the energy correlator as defined in \ZhiboedovOPA. We will use the formalism of \CostaMG\  and restrict our discussion on operators which are symmetric and traceless tensors. The three point function we are interested in is
\eqn\threepoint{
\la {\cal O}_{j}^{\dagger} (x_1, \eps^*) T (x_2, \bar n)  {\cal O}_{j}(x_3, \eps)\ra = { \sum \alpha_{i} V_1^{v_1} V_2^{v_2} V_3^{v_3} H_{12}^{h_{12}} H_{13}^{h_{13}} H_{23}^{h_{23}} \over x_{12}^{d+2} x_{13}^{2 \bar \tau - (d+2)} x_{23}^{ d + 2}}\,,
}
where $\bar \tau = \Delta + j$ and $\bar n = (1, - \vec n)$ and the exponents $v_i$, $h_{ij}$ obey the following constraints
\eqn\spinconstraint{\eqalign{
v_1 + h_{12} + h_{13} &= j, \cr
v_2 + h_{12} + h_{23} &= 2, \cr
v_3 + h_{13} + h_{23} &= j. 
}}
So different structures are labelled by the $\{ h_{12}, h_{23}, h_{13} \}$. Of course, on top of these constraints one should impose the conservation condition or - possibly - permutation symmetry. For our argument imposing those is not necessary.

The method we are using is the one of \BelitskyXXA. The relevant notation is introduced in appendix E. We start by expressing the energy correlator one point function using \efdef\ and \threepoint. We then take the limit  
for the stress tensor approaching null infinity with the help of appendix F. The result can be expressed as follows
\eqn\resultofLIM{\eqalign{
 { \sum \alpha_{\{h_{12}, h_{13}, h_{23} \}} \hat V_1^{v_1} \hat V_2^{v_2} \hat V_3^{v_3} \hat H_{12}^{h_{12}} \hat H_{13}^{h_{13}} \hat H_{23}^{h_{23}} \over (x_{21}.n)^{{d+2 \over 2} } (x_{13}^2)^{ \bar \tau - {(d+2) \over 2} }  (x_{23}.n)^{{d+2 \over 2} }}
}}
where we introduced
\eqn\reducedCONF{\eqalign{
\hat V_1 &= -{x_{13}.\eps^* x_{12}. n - \eps^* . n {x_{13}^2 \over 2} \over x_{23}. n},~\quad\hat V_2 =  {x_{13}.n \over x_{13}^2},~\quad\hat V_3 =- {x_{13}.\eps  \, x_{23}. n  -   \eps.n \,  {x_{13}^2 \over 2}  \over   \ x_{12}. n}, \cr
\hat H_{12} &=-\eps^*.n ,~\qquad \hat H_{13} =\eps^* . \eps \ x_{13}^2 - 2 \ x_{13}.\eps^* x_{13}.\eps,~\qquad \hat H_{23} =-\eps.n .
}}
Setting $\eps . n = 0$ leads to further simplifications.
The three-point function then reduces to
\eqn\corrlim{\eqalign{
\sum_{h_{13}=0}^{j} \alpha_{ \{0, h_{13}, 0 \}}{ \left( \eps^* . x_{13} \eps . x_{13} \right)^{j - h_{13} }  \over x_{12}.n^{{d+2 \over 2}} x_{23}.n^{{d+2 \over 2}} }  {\hat H_{13}^{h_{13}} (x_{13}.n)^2 \over x_{13}^{2 \bar \tau - (d-2) } } \, .
}}

Next, we integrate over the position of the detector $\int_{- \infty}^{\infty} d (x_2. n)$. This boils down to the replacement $(x_{12}.n)^{- {d+2 \over 2}} (x_{23}.n)^{-{d+2 \over 2}} \to (x_{13} . n)^{-(d+1)}$ in the formula above (see appendix F for the precise formula). 
Notice that after this replacement the dimensionality of the object \corrlim\ is $(1+ 2 \Delta)$ as it should be for a correlator which measures energy.\foot{The $2 \Delta$ piece cancels when we divide by the two-point function, which is given by ${H_{13}^{j} \over (x_{13}^2)^{\bar \tau}}$.} 

The final step in the computation of the energy correlator is the Fourier transform, which implements the insertion of an operator with a given momentum. This leads to the following expression for the energy flux one-point function
\eqn\resultofLIMtrans{\eqalign{
\la {\cal E}(n) \ra_{{\cal O.\eps}(k)} &\sim \int_0^{\infty} d s \ s^d \int d^{d} x_{13} e^{-i (k - s n) . x_{13}} \ { \sum_{h_{13}=0}^{j} \alpha_{\{0, h_{13}, 0 \}} (\eps^*.x_{13} \eps.x_{13})^{j - h_{13}} (x_{13}.n)^{2} \hat H_{13}^{h_{13}} \over (x_{13}^2)^{\bar \tau - {d-2 \over 2}}} \geq 0 ,
}}
where we ignored an overall positive constant.
Recall that in the formula above the propagator is the Wightman one and the integral has non-zero support only for $(k-s n)$ time-like and having positive energy. We will not need to compute this integral explicitly.

\subsec{A Computation on the DIS Side}

Let us repeat the computation on the DIS side. We start with the analysis of the OPE.
The relevant formula is the following \CostaMG\
\eqn\OPEcontr{
{\cal O}(\eps^*, x_{13}) {\cal O}(\eps, 0)  \sim {\cal O}(0, \pa_{z}) t(x_{13},\eps^*,z,\eps) x_{13}^{-\left(\Delta_1 + \Delta_3 - \Delta_2 +s_1 + s_2 + s_3 \right)}
}
The polynomial $t(x_{13},\eps^*,z,\eps)$ is fixed by the three point function to be

\eqn\OPEcontr{\eqalign{
t(x_{13},\eps^*,z,\eps) &= \sum \alpha_{\{h_{12}, h_{13}, h_{23} \}} (x_{13}^2)^{v_2+ h_{12}+h_{23}} (-1)^{v_1+v_3} \quad \times \cr 
& \times (\eps^*.x_{13})^{v_1}({x_{13}.z \over x_{13}^2})^{v_2} (\eps.x_{13})^{v_3} (\eps^*.z)^{h_{12}} \hat H_{13}^{h_{13}} (z. \eps)^{h_{23}}\,.
}}
This leading contribution to the sum rules come from the term $h_{12} = h_{23} = 0$ as explained before.\foot{Effectively, this is equivalent to setting $\eps^* . P = \eps . P  = 0$.} 
Moreover, we are interested in the case ${\cal O}(0, \pa_{z}) \to T_{\mu \nu}$. For this case we get
\eqn\OPEcontrDIS{
\la P | {\cal O}(\eps^*, x_{13}) {\cal O}(\eps, 0) | P \ra \sim {\la P | T(0, \pa_{z}) | P \ra \over (x_{13}^2)^{\bar \tau - {d-2 \over 2}}}  \sum_{h_{13}=0}^{j} \alpha_{\{0, h_{13}, 0 \}} (\eps^*.x_{13} \eps.x_{13})^{j - h_{13}} (x_{13}.z )^{2} \hat H_{13}^{h_{13}} .
}
where ``$\sim$'' denotes that we neglected the contribution of all the other operators present in the OPE.
Now it is trivial to act with $\pa_{z}$ which boils down to $z\to p$,

\eqn\OPEcontrDIS{
\la P | {\cal O}(\eps^*, x_{13}) {\cal O}(\eps, 0) | P \ra\sim 2 { \sum_{h_{13} = 0}^{j} \alpha_{\{0, h_{13}, 0 \}} (\eps^*.x_{13} \eps.x_{13})^{j - h_{13}} (x_{13}. P )^{2} \hat H_{13}^{h_{13}} \over (x_{13}^2)^{\bar \tau - {d-2 \over 2}}} \, .
}

Finally, we must take the Fourier transform with respect to $x_{13}$ which, assuming we can trust the dispersion integral, leads to the following constraint

\eqn\DISone{
\int d^4 x e^{- i q x}   { \sum_{h_{13}=0}^{j} \alpha_{\{0, h_{13}, 0 \}} (\eps^*.x_{13} \eps.x_{13})^{s_3 - h_{13}} (x_{13}.P)^{2} \hat H_{13}^{h_{13}} \over (x_{13}^2)^{\bar \tau - {d-2 \over 2}}} \geq 0 ,
}
where $q$ is space-like and $x_{13}^2$ is the usual time-ordered propagator.

\subsec{Relation between Energy Correlators and DIS}

We will now use the results of subsections 4.1 and 4.2 to find a precise relation between the energy correlator and the DIS amplitude. Combining equations \resultofLIMtrans\ and \DISone\ we can express the energy correlator one-point function as follows
\eqn\efdis{\eqalign{
\vev{ {\cal E} (n)}_{{\cal O}.\eps (q)} \sim \int_0^{\infty} d s s^d {\rm Im}_{q^2}[\AA_2(q^2, \eps. q, p.q)]_{q \to q - P ; P \to s n } 
}}
where $\AA_2(q^2,\eps. q,P.q)$ is defined as the term in the full DIS amplitude $\AA(q^2,\eps. q,p.q)$,
\eqn\AAdef{\AA(q^2,\eps. q,P.q)=\int d^dy e^{-iq y} \vev{ P | \TT \left(\OO_j(\eps^\ast, y)\OO_j(\eps, 0) \right) | P}} 
derived from the OPE coefficient of the stress energy tensor operator.  Recall that we consider polarization tensors satisfying $\eps.n =0$.

Eq. \Aj\ allows us to write $\AA_2(q^2, \eps.q, P.q)$ in the following form
\eqn\DISdef{
\AA_2(q^2,\eps. q,p.q)=  (2P.q)^2 \sum_{m=0}^j (q^2-i \varepsilon)^{-d-1+\Delta_j-m} \,\,C_{T,m}\,  \, {(2\eps .q)^m(2\eps^\ast .q)^m (\eps^\ast .\eps)^{j-m}} \,.  
} 
where the Fourier transform has been obtained following the Feynmann -$i\varepsilon$ prescription. The coefficients $C_{T,m}$ are defined as
\eqn\Cmdef{\eqalign{
C_{T,m} &=4^{-\beta} {\pi^{d\over 2}\Gamma[-\Delta_\OO+m+d+1]\over \Gamma[\Delta_\OO+m+1-d/2]} \,a_{T,m}\cr
&\beta=\Delta_\OO+m+1-d\,,
}}
where $a_{T,m}$ denote the constant OPE coefficients of the energy momentum tensor. Observe that the $\Gamma$-function in the denominator of \Cmdef\ is positive definite by unitarity but the one in the numerator is not necessarily positive definite, as discussed in detail in section 3.

The energy flux expectation value in \efdis\ depends on  ${\rm Im}_{q^2} \AA_2(q^2,\eps.q,P.q)$ which is equal to
\eqn\ImAA{\eqalign{{\rm Im}_{q^2} \AA_2(q^2, \eps. q,p.q)&\,\,\, = (2P.q)^2 \theta (q^0)\theta (-q^2) \times \cr
&\times\sum_{m=0}^j \widetilde{C}_{T,m}\,  (2\eps .q)^m(2\eps^\ast .q)^m (\eps^\ast .\eps)^{j-m} (-q^2)^{-d-1+\Delta_j-m}~, 
}}
where $\widetilde{C}_m$ are equal to
\eqn\Ctildem{\widetilde{C}_{T,m}={ 4^{-\beta} \pi^{{d\over 2}+1} \over  \Gamma[\Delta_\OO+m+1-d/2] \Gamma[\Delta_\OO-m-d] } \, a_{T,m}\,.}
Note that the product of $\Gamma$-functions appearing in $\widetilde{C}_{T,m}$ is not positive definite either. For operators of spin-$j$ unitarity implies that $\Delta_\OO-m-d\geq (j-m)-2$ whereas for scalars, {\it {i.e.}}, $j=m=0$, unitarity leads to $\Delta_\OO-d\geq -{d\over 2}-1$. 
Substituting \ImAA\ into \efdis\ yields
\eqn\AAdefa{\eqalign{\vev{\EE(n)}&\sim \sum_{m=0}^j \tilde{C}_m (2\epsilon.q)^m(2\epsilon^\ast .q)^m(\epsilon^\ast.\epsilon)^{j-m}  (2 n.q)^2 \cr
&\times \int_0^\infty ds \, s^{d+2} \theta(q^0-s) \theta(-(q^2-2 s q.n)) \,\left(-(q^2-2 s q.n)\right)^{\Delta_\OO -d-m-1} \cr
&=\sum_{m=0}^j a_m (2\epsilon.q)^m(2\epsilon^\ast .q)^m(\epsilon^\ast.\epsilon)^{j-m}  (2 n.q)^2 \theta(q^0) \theta(-q^2) (-q^2)^{\Delta_\OO-d-m-1} \left(q^2\over 2 q.n\right)^{d+3}\times \cr
&\times {\Gamma(d+3) \over \Gamma(\Delta_\OO-m+3)\Gamma[\Delta_\OO+m+1-d/2]}
}}
Positivity of \AAdefa\ is equivalent to $(j+1)$ positivity relations, one for each value of $m$, obtained by appropriately choosing the polarization tensors $(\epsilon^\ast, \epsilon)$. In other words,
\eqn\posee{\vev{\EE(n)}\geq 0\,\,\Leftrightarrow \,\, a_m {\Gamma(d+3) \over \Gamma(\Delta_\OO-m+3)\Gamma[\Delta_\OO+m+1-d/2]} \geq 0\,.}
Notice, that the $\Gamma$-functions in \posee\ are now positive definite by unitarity, {\it {i.e.}},
\eqn\gammapd{\Delta_\OO\geq d-2+j\geq d-2+m\geq m-3 ,\quad \Delta_\OO\geq d-2+j\geq d/2-m-1}
allowing us to write
\eqn\equivl{\vev{\EE(n)}\geq 0\,\,\Leftrightarrow \,\, a_m\geq 0, \quad m=0,1,\cdots,j\,.}

Eq. \equivl\ establishes the equivalence between the constraints obtained from DIS and those derived from the positivity of the energy correlators for a certain class of transverse polarizations. This class of polarizations, exhausts all possible choices for conserved currents. A generic operator, however, may also have longitudinal polarizations. To examine in detail what happens for generic operators, we consider below the case of a non-conserved spin one current.

\subsec{Non-Conserved Spin One Current}

Let us consider the three-point function which involves two operators of spin one, with a generic twist. Its general form is \ZhiboedovBM
\eqn\generalthreepoint{\eqalign{
\la {\cal O}(x_1, z_1)& T(x_2, z_2) {\cal O}(x_3, z_3) \ra = {1 \over x_{12}^{d+2} x_{13}^{2 \bar \tau - (d+2)} x_{23}^{ d + 2}} \left\{ a_1 V_2^2 H_{13}  \right. \cr
&\left. + {(d-2)^2 a_2 \over 2} \left( V_2^2 V_1 V_3 - V_2 {2V_1 H_{23} + 2 V_3 H_{12} + V_2 H_{13} \over d-2} + {2 H_{12} H_{23} \over (d-2)^2} \right)\right. \cr
&\left. - 2 (d-1) a_3 \left(V_2^2 V_1  V_3 + {V_2 \over 2} [ V_{2} H_{13} + V_1 H_{23} + V_3 H_{12} ]\right) \right\}.
}}
Here $(z_1,z_2,z_3)$ denote the correponding polarization tensors. The coefficients $a_2$ and $a_3$ are proportional to the structures $\la J T J\ra$ generated in the theory of a free boson and free fermion respectively for conserved spin one current $J$.
We can compute the energy correlator as explained above. Stress tensor Ward identities relate this three-point function to the two-point function $\vev{\OO(z_1,x_1) \OO(z_3,x_3)}$ \OsbornCR. This can be translated into a relation between the parameters $(a_1,a_2,a_3)$ appearing in \generalthreepoint. To find this relation, we require instead that the two-point function is correctly reproduced after integrating the energy flux correlator over the position of the detector. This leads to the following constraint
\eqn\constraintTWO{
a_1 =- (\Delta - d + 1) \left(a_2 + a_3 \right) .
}
Notice that for $\Delta = d -1$, which corresponds to the case of a conserved current, $a_1 = 0$ as it should (in this case only two independent structures are expected to appear). Reflection positivity of the two-point function then yields
\eqn\postwop{
\la {\cal O} {\cal O}\ra \sim a_2 + a_3 > 0  \ .
}

Computing the energy flux correlator after imposing \constraintTWO, and requiring it to be positive-definite, results in the following two conditions 
\eqn\resultspinone{\eqalign{
a_2 &< 0, ~~~ a_3 \geq - {a_2 \over (\Delta - d+ 1)} { (\Delta + 1) (2 \Delta + (d-2)(d-1) ) \over 2 \Delta + d - 2} \  , \cr
a_2 &\geq 0, ~~~ a_3 > - {2 (\Delta - d + 1) (\Delta - 1) \over 2 (\Delta - d +1) (\Delta - 1) + d \Delta}  a_2  \ .
}}
We should stress here, that in deriving \resultspinone\ we did not require $\eps.n=0$.
Notice that when $\Delta \to d-1$ the solution which corresponds to the first line of \resultspinone\ disappears, whereas the second line approaches the bounds of conserved currents \HofmanAR. Indeed, for $\Delta = d - 1$ we recover the usual $a_2 \geq 0$ and $a_3 \geq 0$ conditions. 

On the other hand, we can compute the DIS bound or, equivalently, restrict our consideration to $\eps. n = 0 $ in the energy correlator computation. The result is the same for $a_2 \geq 0$ but for the other case we get
\eqn\resultspinoneDIS{\eqalign{
a_2 &< 0, ~~~ a_3 > - {a_2 \over  (\Delta - d+ 1)} { 4 (\Delta - 1) + (d-4)(d-2)  \over 4} \  .
}}
It is easy to see that the bounds derived from the positivity of the energy correlator are stronger than those obtained from DIS, for any $\Delta > d-1$.

\subsec{Non-Conserved Spin Two Current}

Similarly, we computed the energy correlator for a generic non-conserved spin two current. There are six different structures that appear in the three-point function. Matching to the two-point function after integrating over the position of the detector fixes one of the constants in terms of the others. We again find that the constraints from the energy correlator are stronger than the ones from DIS for non-conserved spin-two operator. In the limit $\Delta \to d$ the constraints derived from DIS become equivalent to those required by the positivity of the energy correlator, as predicted by the general argument above.

\newsec{DIS in a CFT}

A consistent unitary CFT should produce correlation functions that are reflection positive. As is often the case, it is easier to analyze the constraints following from reflection positivity in Lorentzian signature. Obviously, these constraints should hold independently of whether or not the CFT admits an RG flow to a gapped phase. In this section we reformulate the sum rules studied in the previous sections purely in the CFT language. Instead of a proton we consider the state $\ket{P}$ defined as follows
\eqn\Pdef{\ket{P}\equiv \int d^d y e^{i P y} \OO(y)\ket{0}}
where $\OO(x)$ is an arbitrary, scalar operator.

The expectation value of the stress energy tensor on the state $\ket{P}$ is determined by Lorentz invariance up to two numbers,
\eqn\vevT{ \vev{P|T_{\mu\nu}|P} =c_1 P_\mu P_\nu+ c_2 \eta_{\mu\nu} P^2} 
where $c_1,c_2$ are some dimensionful coefficients. Conformal invariance allows us to further express $c_2$ in terms of $c_1$. We will not consider the second term on the right hand side of \vevT\ since it belongs to the so called ``trace" terms, whose contribution to the OPE is negligible for large, spacelike momentum. Instead we show in appendix G that $c_1$, up to an overall divergent term, is positive-definite. The divergence can be easily regularized; for example, we can imagine making the norm finite by considering $e^{- {y_0^2 + \vec y^2 \over \sigma^2}}$ for the wave function.

We next consider the DIS amplitude defined as follows
\eqn\DIScft{\eqalign{
&\AA(q^2,\eps. q,P.q)=\int d^dy e^{-iq y} \vev{P | \TT \left(\OO_j(\eps^\ast, y)\OO_j(\eps, 0) \right)| P }_{conn} \cr
&=\int d^dy e^{-iq y} \biggl( \vev{P | \TT \left(\OO_j(\eps^\ast, y)\OO_j(\eps, 0) \right)| P } -   \vev{P | P }  \vev{0 | \TT \left(\OO_j(\eps^\ast, y)\OO_j(\eps, 0) \right)| 0 } \biggr)  \ ,
}}
where operators are ordered as written and $\TT(...)$ stands for time ordering.

The imaginary part of \DIScft\ is positive definite. To see this, recall that the imaginary part of the full correlator is given by the positive-definite Wightman function, and that the imaginary part of the disconnected piece is independent of $x={q^2\over 2 P.q}$ and vanishes for space-like $q^2$.

Let us recall the analytic structure of $\AA(q^2,\eps. q,P.q)$. It has discontinuities for $(P+q)^2<0$ and $(P-q)^2 <0$. These can be rewritten as
\eqn\cond{
- {1 \over 1+{P^2 \over q^2}} \leq x \leq {1 \over 1+{P^2 \over q^2}} \ , 
}
where $x\equiv {q^2\over 2 P.q}$.
We can then proceed as before. We have to assume a certain behavior at infinity to use the dispersion relations but otherwise all the formulae are identical to the ones in the previous sections. Formulated in this language, $B_s^*$ from sections 2 and 3 are simply proportional to the corresponding three-point couplings. This is why~\constraint\ follows.

\newsec{Conclusions}

In this paper we considered the DIS experiment in a unitary CFT. The basic object under consideration is the scattering amplitude 
\Adefgenj . Using it one can write the standard sum rules \ct\ which relate the OPE data to the integrated positive-definite cross section. 

An interesting case to consider is the graviton DIS in a CFT which flows to a gapped phase. In this case the structure of the amplitude is given by \AT\ and the positivity of the cross section leads to the constraints \CT\ which are the well-known Hofman-Maldacena constraints. More general constraints exist in each even spin sector \newbounds. These can be therefore viewed as generalized Hofman-Maldacena constraints.

We studied the general DIS experiment with some spinning external operators and elucidated the relation between the bounds produced by DIS and energy correlator considerations. Our first conclusion is that the $s=2$ DIS bounds are equivalent to the energy flux constraints computed for a subclass of polarization tensors. This follows from the relation \efdis\ which is the result of an explicit computation. 

Considering the DIS experiment which involves non-conserved spin-one and spin-two currents, we found that, generically, the constraints obtained from the energy flux positivity are stronger than those coming from the DIS $s=2$ sum rule, as explained in subsection 3.4. The difference between the two methods disappears in the limit when the operators become conserved. Understanding better the origin of this difference is an important open problem. 

Finally, we reformulated the DIS experiment purely in CFT terms. The role of the DIS amplitude is played by the four-point correlation function \DIScft, with the particular ordering of operators and a special choice of external wave functions. Positivity of the cross section translates to positivity of the norm and the usual problem with the dominance of the unit operator does not appear because of the choice of ordering and kinematics. 

In writing the sum rules we assumed a certain behavior of the amplitude in the Regge limit. We expect that this behavior depends both on the details of the theory and on the properties of the operators involved. This can be easily seen to be the case for the free scalar theory (see appendix D). In the bulk of the paper we simply assumed that we can write the sum rules and derived the consequences. However, for the graviton DIS, it is legitimate to assume that all the sum rules for $s\geq 2$ converge. This produces various results that are satisfied in all the examples known to us.

Our analysis indicates that there are infinitely more constraints on the three-point function of spinning operators than have been known before. It would be very interesting to see what can be learned about them using other methods. These include integrability \BassoZOA ; numerical \RattazziPE\ and analytic bootstrap \LiITL ; and casuality \HartmanLFA . 

There are several directions in which our analysis can be generalized. One direction involves considering operators in generic representations of the Lorentz group both for the probe operator and for the target, including the parity odd structures in $d=3$. One may also consider the odd spin sum rules. In this case the OPE data for the minimal twist odd spin operators is related to the difference of cross sections of the type $\sigma_{p n} - \sigma_{p \bar n}$ where $\bar n$ refers to an anti-particle. This difference is not known to be positive-definite. The recent conjecture for bounds on the $\la J J J\ra$ three-point coupling put forward in \LiITL, together with the convergence of the $s=1$ sum rule would imply the sign-definiteness of (roughly) $\int_0^{\infty} d x \left(\sigma_{p n} - \sigma_{p \bar n} \right) \geq 0$. It would be interesting to investigate this further.

Another interesting open avenue is bounding the non-integrated expectation value of the stress tensor in a given state and deriving the consequences for the OPE. In a classical theory the expectation value of the stress energy tensor is non-negative. In a quantum theory however the expectation value can be locally negative but as reviewed in this paper, the integrated over time expectation value is expected to be non-negative. A more refined version of this statement is that there are bounds on how negative the local expectation value of the stress tensor in a given state could be \refs{\FordID,\FordQV} . This was recently discussed in \refs{\BoussoMNA,\BoussoWCA}\ and in \FarnsworthHUM\ where bounds on the three-point function of the stress tensor in a unitary 4d CFT were obtained. It would be interesting to understand if these bounds could be strengthened and what constraints on the spectrum and the three-point functions of the CFT they imply.

One of the puzzling features of AdS/CFT is the emergence of locality on the sub-AdS scale. It is believed \HeemskerkPN\ that CFTs with large central charge  $N$ and large gap in the spectrum of higher spin currents $\Delta_{gap} \gg 1$, are described at low energies by Einstein's theory in $AdS_{d+1}$ with all higher-derivative corrections suppressed by the gap $\Delta_{gap}^{- 1}$. Proving this using purely CFT methods seems to be a necessary and important step in our understanding of the AdS/CFT correspondence and more generally quantum gravity. An even more ambitious goal is to show that every theory with such properties is a string theory. We hope that methods developed in this paper could be useful to make progress in this direction.

In \CamanhoAPA\ it was shown, using bulk arguments, that the picture described above follows from causality in the case of the simplest possible observable, namely the graviton self-coupling. Showing this for all correlation functions and using purely CFT methods is still an open problem.

\vskip 1cm

\noindent {\bf Acknowledgments:}

We would like to thank  Tom Hartman and Adam Schwimmer for useful discussions. 
Z.K. is supported by the ERC STG grant 335182, by
the Israel Science Foundation under grant 884/11, by
the United States-Israel Bi-national Science Foundation
(BSF) under grant 2010/629, by the Israel Science Foundation
center for excellence grant (grant no.1989/14)
and also by the I-CORE Program of the Planning and
Budgeting Committee. A.P. is supported in part by a VIDI grant from NWO, the Netherlands Organization for Scientific Research. M.K. is supported in part by a Marie-Curie fellowship with project no 203972 within the European Research and Innovation Programme EU H2020/2014-2020. A.Z. is supported in part by U.S. Department of Energy grant de-sc0007870 and by the Deutsche Forschungsgemeinschaft (DFG grant Ot 527/2-1). We would like to thank CERN, KITP and the Lorentz Center for hospitality at various stages of this project. MK and AP are also grateful to the Weizmann Institute of Science and to the Simons Center for Geometry and Physics for hospitality.

\appendix{A}{The relation between $a_{s,m}^*$ and $\hat{a}_{s,m}^*$ in the $TT\sim T$ OPE.}

In this appendix we present the relation between the Fourier transformed OPE coefficients $\hat{a}_{s,i}^*$ and the coefficients $a_{s,i}^*$ which characterize the OPE in position space, as defined in \fdef.
\eqn\avshata{\eqalign{a_{s,0}^*&=i^s \left(\hat{a}_{s,0}^*+{\hat{a}_{s,1}^*\over 2 d-\tau_s^*}+{2 \hat{a}_{s,2}^*\over (2 d-\tau_s^*)(2 d-\tau_s^*+2)}\right) \ , \cr
a_{s,1}^*&=- i^{s} 4\left(\hat{a}_{s,1}^*+{4 \hat{a}_{s,2}^*\over 2 d-\tau_s^*+2}\right) \ , \cr
a_{s,2}^*&=i^s \hat{a}_{s,2}^*  \ .
}}

\appendix{B}{DIS for the stress-energy tensor operator}

\noindent Here we explicitly evaluate the $a_{T,m}$ and derive \CT. Our starting point is eq.(6.38) of \OsbornCR. Requiring that $\eps^2=(\eps^\ast)^2=\eps.P=\eps^\ast.P=0$ and neglecting ``trace" terms, leads us to consider only the following terms from eq.(6.38) of \OsbornCR:
\eqn\hata{ \eqalign{ \hat{A}^1_{\mu\nu\sigma\rho\alpha\beta}(y) C_T \longrightarrow &{d-2\over d+2} (4a+2b-c) H^1_{\alpha\beta\mu\nu\sigma\rho}(y) +{da+b-c\over d} H^2_{\alpha\beta\mu\nu\rho\sigma}(y)+\cr
&+{2da+2b-c\over d(d-2)} H^3_{\alpha\beta\mu\nu\sigma\rho}(y) }}
Evaluating the Fourier transform together with the appropriate contractions yields:
\eqn\terma{\eqalign{  \int d^d y e^{-i q y} \eps^{\ast,\mu} \eps^{\ast,\nu}  &H^1_{\alpha\beta\mu\nu\sigma\rho} \eps^\sigma \eps^\rho P^\alpha P^\beta =  \cr
& -{1\over 8} {\pi^{d/2}\over \Gamma(d/2+1)}  {(2\eps^\ast.q)^2 (2\eps.q)^2\over (q^2)^2} (2 P.q)^2  q^{-2} + \cr
&+{1\over 2} {\pi^{d/2}\over \Gamma(d/2+1)}  {(\eps^\ast.\eps)(2\eps^\ast.q)(2\eps.q)\over q^2}  (2 P.q)^2 q^{-2} -\cr
&-{1\over 2} {\pi^{d/2}\over \Gamma(d/2+1)} (\eps^\ast.\eps)^2 (2P.q)^2 q^{-2}
}}
\eqn\termbc{\eqalign{\int d^d y e^{-i q y} \eps^{\ast,\mu} \eps^{\ast,\nu}  H^2_{\alpha\beta\mu\nu\sigma\rho} \eps^\sigma \eps^\rho P^\alpha P^\beta&= {\pi^{d/2}\over \Gamma(d/2)} {(\eps^\ast.\eps)(2\eps^\ast.q)(2\eps.q)\over q^2}  (2 P.q)^2 q^{-2} \cr
\int d^d y e^{-i q y} \eps^{\ast,\mu} \eps^{\ast,\nu}  H^3_{\alpha\beta\mu\nu\sigma\rho} \eps^\sigma \eps^\rho P^\alpha P^\beta&=-2 {\pi^{d/2}\over\Gamma(d/2-1)} (\eps^\ast.\eps)^2 (2P.q)^2 q^{-2} 
}}
Collecting the appropriate factors from \terma-\termbc\ leads to \CT.

\appendix{C}{Scalar DIS}

Here we consider the simplest example of a DIS experiment, where the external operator is a scalar. 
Following \OsbornCR\ we consider
\eqn\threepointb{\vev{T_{\mu\nu}(x_1)\OO(x_2)\OO(x_3)}={1\over x_{12}^d x_{23}^{2\Delta-d} x_{31}^d} \II_{\mu\nu,\sigma\rho(x_{13})} t_{\sigma\rho}(X_{12})}
where 
\eqn\twopointx{\eqalign{&\langle \OO(x)\OO(0)\rangle ={N\over x^{2\Delta}}\cr
&\langle T_{\mu\nu}(x)T_{\rho\sigma}(0)\rangle=\CC_{T}{\II_{\mu\nu,\rho\sigma}(x)\over x^{2d}}\cr
&\II_{\mu\nu,\rho\sigma}={1\over 2}\left(I_{\mu\sigma}(x) I_{\nu\rho}(x)+I_{\mu\rho}(x)I_{\nu\sigma}(x)\right)-{\eta_{\mu\nu}\eta_{\rho\sigma}\over d}\cr
&I_{\mu\nu}=\eta_{\mu\nu}-2{x_\mu x_\nu\over x^2}
}}
and 
\eqn\tmunu{t_{\mu\nu}=a\left(\hat{X}_\mu \hat{X}_\nu-{1\over d}\eta_{\mu\nu}\right)\qquad \hat{X}_\mu={X_\mu\over \sqrt{X^2}}}
and $X_{ij}$ is defined as
\eqn\Xdef{X_{ij}=-X_{ji}={x_{ik}\over x_{ik}^2} -{x_{jk}\over x_{jk}^2},\qquad X_{ij}^2={x_{ij}^2\over x_{ik}^2 x_{jk}^2}\qquad i=1,2,3,\,\, i\neq j,\,j\neq k,\,i\neq k\quad .}
The three point function \threepointb\ in the limit $y^\mu\equiv (x_2-x_3)^\mu\rightarrow 0$ yields
\eqn\opecoef{\eqalign{\vev{T_{\mu\nu}(x_1)\OO(x_2)\OO(x_3)} &\simeq_{y\sim 0} 
{\II_{\mu\nu,\rho\sigma}(x_{13})\over x_{13}^{2d}} (-)^d {t_{\rho\sigma}(y)\over y^{2\Delta-d} }\simeq\cr
&\simeq C_{\rho\sigma}(y) C_T {\II_{\mu\nu,\rho\sigma}(x_{13})\over x_{13}^{2d}} \quad\Rightarrow\quad C_{\rho\sigma}(y) C_T=(-)^d {t_{\rho\sigma}(y)\over s^{2\Delta-d}}\,,
}}
leading to
\eqn\opeOOT{\OO(y)\OO(0)\sim \cdots + {C_{\mu\nu}(y)} T^{\mu\nu}(0)+\cdots \,,}
where the dots respresent the contributions of other operators in the OPE.
We therefore express $C_{\mu\nu}(y)$ as follows
\eqn\Cdef{C_{\mu\nu}(y)={a\over C_T} \left[{1\over 4(\Delta-{d\over 2}-1)(\Delta-{d\over 2})}\p_\mu\p_\nu {1\over y^{\Delta-{d\over 2}-1}} -\eta_{\mu\nu} {\Delta-d\over d(\Delta-{d\over 2})} {1\over y^{\Delta-{d\over 2}}}\right]}
$C_{\mu\nu}$ does not contribute to conformal Ward Identities as can be immediately seen from eqs (13.a) and (13.b) of \OsbornCR. This is to be contrasted with the ope coefficient $\hat{C}_{\mu\nu}$ in the ope $\OO(s) T_{\mu\nu}(0)\sim \hat{C}_{\mu\nu}(s) \OO(0)$.
In the latter case, the conformal Ward Identities relate the three point function coefficient $a$ with the coefficient $N$ of the two point function $\vev{\OO(x)\OO(0)}$ in \twopointx\ as follows (see eq. (6.20) of \OsbornCR) :
\eqn\arel{a=-{d N \Delta\over (d-1) S_d} }
For a unitary CFT, \arel\ implies that $a\leq 0$. 

To obtain the contribution of the stress energy tensor in the scalar DIS amplitude $\AA(x,q^2)=\int d^dy e^{-i q y}\langle P|\OO(y)\OO(0) |P\rangle$, we simply need to take the Fourier transform of the ope coefficient \Cdef, taking into acount the expectation value of the stress energy tensor and disregarding the ``trace" terms. Explicitly we have that
\eqn\scalarDIS{\eqalign{\AA(x,q^2)&= \cdots-B_{T} {a\over C_T} {1\over 4(\Delta-{d\over 2}-1)(\Delta-{d\over 2})} (P.q)^2\int d^dy e^{-i q y} y^{-\Delta+d/2+1}+\cdots \cr
&=\cdots-B_T {a\over C_T} (2 q.P)^2 {1\over 16} {\pi^{d/2+1} \Gamma(d-\Delta +1) \over\pi\Gamma(\Delta-d/2+1)} (q^2/4-i\varepsilon)^{\Delta-d-1}+\cdots\cr
&=\cdots -B_T {a\over C_T}  {\pi^{d/2+1} \Gamma(d-\Delta +1) \over\pi\Gamma(\Delta-d/2+1)} (q^2/4-i\varepsilon)^{\Delta-d+1} x^{-2}+\cdots
}}
Notice that the sign of the term in \scalarDIS\ is equal to the sign of the $\Gamma$--function in the numerator. This is because, $a<0$ from the Ward Identities and eq.\arel, $B_T=1$, and unitarity requires that $C_T>0$ and $\Delta-d/2+1>0$. As long as $\Delta<d+1$, the $\Gamma$-function is positive definite and the positivity constraints from DIS trivially satisfied. On the other hand, when $\Delta>d+1$
the Fourier integral is divergent and the moments not well-defined. Naively applying the DIS positivity relations \uniA\ for this case, would lead to inconsistencies due to the periodically alternating sign of the $\Gamma$-function.

\appendix{D}{DIS in the free field theory}

\noindent Let us consider a free massless scalar $\phi(x)$ and the DIS amplitude for $\phi^{n+1}(x)$,
\eqn\amplitude{\eqalign{
{\cal A}&(P,q) = \int d^d x e^{- i q x} \la P | \phi^{n+1}(x) \phi^{n+1}(0) | P \ra = \int d^d x { e^{- i (q - P) x} +  e^{- i (q + P) x} \over (x^{2} + i \varepsilon)^{ n {d - 2 \over 2 } } } \cr
&\sim \left[ \left((q-P)^2 - i \varepsilon \right)^{n {d-2 \over 2} - {d \over 2}} + \left((q+P)^2 - i \varepsilon \right)^{n {d-2 \over 2} - {d \over 2}}\right]  {\Gamma({d \over 2} - n {d-2 \over 2} ) \over \Gamma(n {d-2 \over 2}  )} \cr
&= \left( q^2 \right)^{n {d-2 \over 2}  - {d \over 2}} \left[ \left( 1 - {1 \over x} - {M^2 \over q^2} - i \varepsilon \right)^{n {d-2 \over 2}  - {d \over 2}} + \left( 1 + {1 \over x} - {M^2 \over q^2} - i \varepsilon \right)^{n {d-2 \over 2}  - {d \over 2}} \right] \times \cr
&\qquad\qquad\qquad\times {\Gamma({d \over 2} - n {d-2 \over 2} ) \over \Gamma(n {d-2 \over 2} )}\,.
}}
We see explicitly what is happening in this example. The amplitude has an imaginary part exactly where expected; for $- (1 - {M^2 \over q^2} )^{-1} \leq x \leq (1 - {M^2 \over q^2} )^{-1}$. Moreover, the imaginary part is positive-definite. 

To write the dispersion relations we have to explore the behavior at $x \to 0$. The amplitude behaves as
\eqn\amplbeh{
{\cal A}(P,q) \sim x^{- \left(n{d-2 \over 2} - {d \over 2} \right)},
}
for the sum rule to converge we thus get
\eqn\sumruleconvergence{
s_0 > n {d-2 \over 2}  - {d \over 2} = \Delta - d + 1.
}

The $\Gamma$ function which appears in the Fourier transform of the OPE that becomes negative has the argument
\eqn\argument{
\Gamma \left( s_0 - \left( \Delta - d + 1 - {\tau^* - (d - 2) \over 2} \right) \right).
}
For the case at hand $\tau^* = d-2$ and we see that the convergence of the sum rule goes in parallel with the positivity of the $\Gamma$-function .

\appendix{E}{Kinematics of the three-point functions}

Here we collect some of the notation that we used in the bulk of the paper.
\eqn\spinconstraint{\eqalign{
n&=(1, \vec n),~~~ \bar n = (-1, \vec n), ~x.n = -t + \vec x . \vec n, \cr
x_{i j}^2 &= - (t_i - t_j)^2 + (\vec x_i - \vec x_j)^2,~ x_{\pm;i j}^2 = - (t_i - t_j  \pm i \eps)^2 + (\vec x_i - \vec x_j)^2, \cr
- 2 P_i . P_j &= x_{i j}^2,~ Z_i . Z_j = z_i . z_j ,~P_i . Z_j = x_{ij} . z_j .
}}
The conformal covariants, as defined in \CostaMG, are given by
\eqn\spinconstraint{\eqalign{
V_1 &= V_{1,23} = {Z_1 . P_2 P_1 . P_3 - Z_1 . P_3 P_1.P_2 \over P_2 . P_3} = - {x_{12}.z_1 x_{13}^2 - x_{13}.z_1 x_{12}^2 \over x_{23}^2}, \cr
V_2 &= V_{2,31} = {Z_2 . P_3 P_2 . P_1 - Z_2 . P_1 P_2.P_3 \over P_1 . P_3} =-  {x_{23}.z_2 x_{12}^2 + x_{12}.z_2 x_{23}^2 \over x_{13}^2}, \cr
V_3 &= V_{3,12} = {Z_3 . P_1 P_3 . P_2 - Z_3 . P_2 P_3.P_1 \over P_1 . P_2} =- { x_{23}.z_3 x_{13}^2 -x_{13}.z_3 x_{23}^2 \over x_{12}^2}, \cr
H_{12} &= - 2 \left( Z_1 . Z_2 P_1 . P_2 - Z_1 . P_2 Z_2. P_1 \right) = z_1 . z_2 x_{12}^2 - 2 x_{12}.z_1 x_{12}.z_2, \cr
H_{13} &= - 2 \left( Z_1 . Z_3 P_1 . P_3 - Z_1 . P_3 Z_3. P_1 \right) = z_1 . z_3 x_{13}^2 - 2 x_{13}.z_1 x_{13}.z_3, \cr
H_{23} &= - 2 \left( Z_2 . Z_3 P_2 . P_3 - Z_2 . P_3 Z_3. P_2 \right) = z_2 . z_3 x_{23}^2 - 2 x_{23}.z_2 x_{23}.z_3 . \cr
}}

\appendix{F}{Computing the energy correlator}

To compute the energy correlator we have to first take the limit $\lim_{x_2.\bar n \to \infty} (x_2 . \bar n)^{d-2}$ of \threepoint. Here we write down some useful formulae which allow us to analyze the limit easily\foot{$x_2 ={x_2 .n \over 2} \bar n + {x_2 . \bar n \over 2} n $.}

\eqn\usefullimit{\eqalign{
\lim_{x_2.\bar n \to \infty} x_{12}^2 &= - x_2.\bar n \ x_{12}. n ,~ \lim_{x_2.\bar n \to \infty} x_{23}^2 = x_2.\bar n \ x_{23}. n , \cr
\lim_{x_2.\bar n \to \infty} x_{12}.z_1 &= - {1 \over 2} x_2.\bar n \ z_1.n ,~\lim_{x_2.\bar n \to \infty} x_{23}.z_3 = {1 \over 2} x_2.\bar n  z_3.n, \cr
\lim_{x_2.\bar n \to \infty} x_{13}.z_1 &= x_{13}.z_1 ,~\lim_{x_2.\bar n \to \infty} x_{13}.z_3 = x_{13}.z_3 , \cr
\lim_{x_2.\bar n \to \infty} x_{23}.z_2 &=x_2.\bar n  ,~ \lim_{x_2.\bar n \to \infty} x_{12}.z_2 = -x_2.\bar n ,
}}
Using \usefullimit, we find that the covariant structures of \spinconstraint\ can be expressed in this limit as follows
\eqn\spinconstraintLIM{\eqalign{
V_1 &\to - {x_{13}.z_1 x_{12}. n - z_1 . n {x_{13}^2 \over 2} \over x_{23}. n},~V_2 \to (x_2.\bar n)^2 {x_{13}.n \over x_{13}^2},~V_3\to - {x_{13}.z_3  x_{23}. n  -   z_3.n   {x_{13}^2 \over 2}  \over   \ x_{12}. n}, \cr
H_{12} &\to   - (x_2.\bar n)^2 z_1.n ,~H_{13} \to z_1 . z_3 x_{13}^2 - 2 x_{13}.z_1 x_{13}.z_3,~H_{23} \to - (x_2.\bar n)^2 z_3.n . \cr
}}
Thus, in the limit we get
\eqn\spinconstraintLIMtwo{\eqalign{
{1 \over (x_{12}^2)^{{d+2 \over 2} } (x_{13}^2)^{ \bar \tau - {(d+2) \over 2} }  (x_{23}^2)^{{d+2 \over 2} }} &\to {1 \over (x_2.\bar n)^{d+2}} {1 \over (x_{21}.n)^{{d+2 \over 2} } (x_{13}^2)^{ \bar \tau - {(d+2) \over 2} }  (x_{23}.n)^{{d+2 \over 2} }} \cr
\sum \alpha_{i} V_1^{v_1} V_2^{v_2} V_3^{v_3} H_{12}^{h_{12}} H_{13}^{h_{13}} H_{23}^{h_{23}} &\to (- (x_2.\bar n)^2)^{v_2 + h_{12} + h_{23} } \sum \alpha_{i} \hat V_1^{v_1} \hat V_2^{v_2} \hat V_3^{v_3} \hat H_{12}^{h_{12}} \hat H_{13}^{h_{13}} \hat H_{23}^{h_{23}}
}}
Notice that due to \spinconstraint, $v_2 + h_{12} + h_{23} = s_2 = 2$, and the correlator has the expected asymptotic behavior ${1 \over (x_2.\bar n)^{d-2}}$. 

Gathering the results above, we conclude that after taking the limit, the correlator reduces to
\eqn\resultofLIMa{\eqalign{
 { \sum \alpha_{\{h_{12}, h_{13}, h_{23} \}} \hat V_1^{v_1} \hat V_2^{v_2} \hat V_3^{v_3} \hat H_{12}^{h_{12}} \hat H_{13}^{h_{13}} \hat H_{23}^{h_{23}} \over (x_{21}.n)^{{d+2 \over 2} } (x_{13}^2)^{ \bar \tau - {(d+2) \over 2} }  (x_{23}.n)^{{d+2 \over 2} }}
}}
where we introduced
\eqn\reducedCONF{\eqalign{
\hat V_1 &= -{x_{13}.z_1 x_{12}. n - z_1 . n {x_{13}^2 \over 2} \over x_{23}. n},~\hat V_2 =  {x_{13}.n \over x_{13}^2},~\hat V_3 =- {x_{13}.z_3  x_{23}. n  -   z_3.n   {x_{13}^2 \over 2}  \over   \ x_{12}. n}, \cr
\hat H_{12} &=-z_1.n ,~\hat H_{13} =z_1 . z_3 x_{13}^2 - 2 x_{13}.z_1 x_{13}.z_3,~\hat H_{23} =-z_3.n .
}}

For the integration over the position of the detector recall the following formula
\eqn\detectorintegral{\eqalign{ 
\int_{- \infty}^{\infty} d (x_2. n) {1 \over (x_{12}.n)^{a}  (x_{23}.n)^{b} } = {2 \pi i  \over (x_{13} . n)^{a+b - 1}} { \Gamma(a+b - 1) \over \Gamma(a) \Gamma(b) } .
}}

\appendix{G}{Positivity of the expectation value of the stress-energy tensor in a CFT}

\noindent Let us define a state $\ket{P}$ as follows:
\eqn\Pdef{\ket{P}\equiv \int d^d x e^{i P x} \OO(x)\ket{0}}
where $\OO(x)$ is an arbitrary, scalar operator.
The expectation value of the stress energy tensor on the state $\ket{P}$ is determined by Lorentz invariance up to two numbers,
\eqn\vevT{ \vev{P|T_{\mu\nu}|P} =c_1 P_\mu P_\nu+ c_2 \eta_{\mu\nu} P^2} 
where $c_1,c_2$ are some dimensionful coefficients. Conformal invariance allows us to further express $c_2$ in terms of $c_1$. Here we will not consider the second term on the right hand side of \vevT\ since it belongs to the so called ``trace" terms, whose contribution in the ope is negligible for large, spacelike momentum. We would like instead to determine explicitly $c_1$ and show that, up to an overall divergent term which can be easily regularized, it is positive definite.

To this end, we consider the following three point function in the CFT
\eqn\sa{\int d^d x d^d y e^{i P x} e^{-i P y}\vev{\OO(y) T_{\mu\nu} n^\mu n^\nu \OO(x)} }
with $n^\mu$ some null vector and $\OO(x)$ the scalar operator of conformal dimension $\Delta$ associated to the state $\ket{P}$. For convenience, will work in light-cone coordinates with 
\eqn\m{ds^2=-dx^{+}dx^{-}+\delta_{ij} dx^i dx^j} 
and choose $n^\mu=(0,1,0,\cdots,0)$ so that we only need to compute $\vev{P|T_{--}|P}$. With this choice we expect that $\vev{P|T_{\mu\nu}|P}=c_1 (P.n)^2={c_1 \over 4} (P^+)^2$, disregarding for the moment possible divergences. 

We start from the general form of the three point function of two scalar operators and the stress energy tensor, as given in (3.1) of \OsbornCR\
\eqn\threea{\vev{\OO(y)T_{\mu\nu}(0) \OO(x)} ={1\over y^{ d} x^d (y-x)^{2\Delta-d} } t_{\mu\nu}(X_{23}) }
where
\eqn\threeb{t_{\mu\nu}(X)=a (\hat{X}_\mu \hat{X}_\nu-{1\over d} \eta_{\mu\nu}), \qquad \hat{X}_{\mu}={X_\mu\over\sqrt{X^2}}\,.}
$X_{ij}$ is defined as follows
\eqn\threec{X_{ij}=-X_{ji}={x_{ik}\over x_{ik}^2} -{x_{jk}\over x^2_{jk}} ,\qquad X_{ij}^2={x_{ij}^2\over x_{ik}^2 x_{jk}^2}\qquad i\neq j,\,j\neq k,\,k\neq i \,.}
Note that the overal coefficient $a$ is completely determined by Ward Identities (see eq. (6.20) of \OsbornCR),
\eqn\adef{a=- {d N \Delta\over (d-1) S_d} <0}
with $N$ the normalization constant of the two-point function of $\OO(x)$ and $S_d$ the volume of the $d$-dimensional sphere.
Using {\threea-\threec} leads to
\eqn\intone{\vev{P|T_{--}|P}={a\over 4} \int d^dx d^d y  {(y^+x^2-x^+y^2)^2e^{iP(x-y)}\over (y^2+i \epsilon y^0)^{{d\over 2}+1} (x^2-i \epsilon x^0)^{d/2+1} \left[(y-x)^2+i\epsilon (y^0-x^0)\right]^{\Delta-d/2+1} }}
where the $-i\epsilon$ presiption is the appropriate one for the Wightman correlator. We split the integral in \intone\ into three separate integrals by expanding the square in the numerator. Each integral is of the form
\eqn\Idef{I_{m,\ell} (P)=\int d^d x d^d y { (y^{+} x^2)^m (x^+ y^2)^\ell e^{i P(x-y)}\over (y^2+i \epsilon y^0)^{{d\over 2}+1} (x^2-i \epsilon x^0)^{d/2+1} \left[(y-x)^2+i\epsilon (y^0-x^0)\right]^{\Delta-d/2+1} } }
where $m,\ell=0,1,2$ and $m+\ell=2$.

First, we express each factor in the denominator as follows
\eqn\dena{\eqalign{x^2-i \epsilon &=(x^+ +i\epsilon )\, \left(-x^-+\sum_{i} {(x^i)^2\over x^+} -i\epsilon\right)\cr
y^2+i \epsilon &=(y^+-i\epsilon )\, \left(-y^-+\sum_{i} {(y^i)^2\over y^+} +i\epsilon\right)\cr
(y-x)^2+i\epsilon&=(y^+-x^+-i\epsilon)\, \left(-y^-+x^-+\sum_{i} {(y^i-x^i)^2\over y^+-x^+} +i\epsilon\right)\,.
}}
Then, we introduce Schwinger parameters 
\eqn\schsi{\eqalign{&\left(-x^-+\sum_{i} {(x^i)^2\over x^+} -i\epsilon\right)^{d/2+1-m}={i^{d/2+1-m} \over \Gamma (d/2+1-m)} \int_0^\infty d{s_1} s_1^{d/2-m}e^{-i s_1\left(-x^-+\sum_{i} {(x^i)^2\over x^+} -i\epsilon\right) }\cr
&\left(-y^-+\sum_{i} {(y^i)^2\over y^+} +i\epsilon\right)^{d/2+1-\ell}={(-i)^{d/2+1-\ell} \over \Gamma (d/2+1-\ell)} \int_0^\infty d{s_2} s_2^{d/2-\ell} e^{i s_2\left(-y^-+\sum_{i} {(y^i)^2\over y^+} +i\epsilon\right) }\cr
&\left(-y^-+x^-+\sum_{i} {(y^i-x^i)^2\over y^+-x^+} +i\epsilon\right)^{\Delta-d/2+1} \!\!\!\! \!\!\!\!\!={(-i)^{\Delta-d/2+1} \over \Gamma (\Delta-d/2+1)} \times\cr
&\qquad\qquad\qquad\qquad\qquad\qquad\qquad\times  \int_0^\infty d{s_3} s_3^{\Delta-d/2} e^{i s_3\left(-y^-+x^-+\sum_{i} {(y^i-x^i)^2\over y^+-x^+} \right) }\,,
}}
and over $(x^-,\, y^-)$ to obtain $(2\pi)^2 \delta(s_1+s_3-p^+/2)\delta(-s_2-s_3+p^+/2)\theta({P^+\over 2}-s_3)$. The integration over $(s_1,s_2)$ then becomes trivial; amounts to setting $s_1=s_2={P^+\over 2}-s_3$ and multiplying with an overall factor of $(1/2)^2$ -- note that the integrations of the $\delta$-fucntions are from zero to infinity. 

In what follows, we find it convenient to keep the variables $s_1,\, s_2$ and perform the relevant substitution later. The next step is to perform the integration over the $(x^i)$, and then over the $(y^i)$. The former reads
\eqn\inttwo{\eqalign{\int \left(\prod_{i=1}^{d-2} dx^i\right) \, {\rm\exp} &\left(i P^i x^i-i s_1 {(x^i)^2\over x^+} +i s_3 {(x^i)^2\over y^+-x^+}-2 i s_3{ y^ix^i\over y^+-x^+} \right)=\cr
&={\pi^{{d-2\over 2}} \over \left(-i A\right)^{{d-2\over 2}}} 
\,\, {\rm\exp} \left\{ -i {\sum_{i}\left(P^i-2 s_3 {y^i\over y^+-x^+}\right)^2 \over 4 A}  \right\}\,
}}
and the latter
\eqn\intthree{\eqalign{\int \left(\prod_i^{d-2} dy^i\right) 
{\rm\exp}\,&\left(-i P^i y^i+i s_2 {(y^i)^2\over y^+}+i s_3 {(y^i)^2\over y^+-x^+} -i {s_3^2\over A} { (y^i)^2\over (y^+-x^+)^2}+i {s_3\over A} P^i {y^i\over y^+-x^+}\right)=\cr
&={\pi^{{d-2\over 2}}\over (-i B)^{{d-2\over 2}}} \, {\rm\exp} \,
\left\{ -i {\sum_i (P^i)^2\left(-1+{s_3\over A (y^+-x^+)}\right)^2\over 4 B}    \right\} 
}}
where 
\eqn\ABdef{A\equiv -{s_1\over x^+}+{s_3\over y^+-x^+} ,\qquad B\equiv {s_2\over y^+}+{s_3\over y^+-x^+}-{s_3^2\over A (y^+-x^+)^2}\,.}
Next follows the integration over the variables $x^+, \, y^+$, i.e.,
\eqn\intfour{\int dx^+ \, dy^+\, {(-A B)^{{2-d\over 2}}\, {\rm\exp}\,
\left\{-i {\overrightarrow{P}^2\over 4}\left({ \left(-1+{s_3\over A (y^+-x^+)}\right)^2  \over B}+{1\over A}\right) -i{P^-\over 2} x^++i {P^-\over 2} y^+ \right\}  \over (x^++i\epsilon)^{d/2+1-\ell-m} (y^+-i\epsilon)^{d/2+1-m-\ell} (y^+-x^+-i\epsilon)^{\Delta-d/2+1}}   \,. }
A little bit of algebra combined with the substitution $s_1=s_2={P^+\over 2}-s_3$ yields
\eqn\algebra{\eqalign{-A B&={P^+\over 2} {({P^+\over 2}- s_3)\over x^+ y^+}\cr
{ \left(-1+{s_3\over A (y^+-x^+)}\right)^2  \over B} +{1\over A}  &={2 (y^+-x^+)\over P^+} }}
which allows us to express \Idef\ as follows
\eqn\intbf {\eqalign{I_{m,\ell}(P)&= C \,\theta\left({P^+\over 2}\right)\int_0^{P^+/2} ds s^{\Delta-d/2} \left( {P^+\over 2}-s\right)^{d/2-m-\ell+1} \times\cr
&\times\int dx^+ dy^+ 
{  e^{-i {P^2\over 2P^+} (y^+-x^+) } \over (x^++i\epsilon)^{2-\ell-m} (y^+-i\epsilon)^{2-\ell-m} (y^+-x^+-i\epsilon)^{\Delta-d/2+1}  }
} }
where
\eqn\Cdef{C\equiv \left({P^+\over 2}\right)^{1-d/2} {\pi^2 i^{-m} (-i)^{-\ell+\Delta-d/2+1} \pi^{d-2} \over \Gamma(d/2+1-m)\Gamma(d/2+1-\ell)\Gamma(\Delta-d/2+1)}\,.}
The integral over $s$ is equal to
\eqn\ints{\eqalign{\!\!\!\!\!\int_0^{P^+/2} ds s^{\Delta-d/2}  &\left( {P^+\over 2}-s\right)^{d/2-m-\ell+1}\!\!\!\!\!\! =
\left( {P^+\over 2}\right)^{\Delta+2-m-\ell} \!\!\!\!\int_0^1 du u^{\Delta-d/2} (1-u)^{d/2-m-\ell+1}= \cr
&\qquad\qquad\,\,=\left( {P^+\over 2}\right)^{\Delta+2-m-\ell} {\Gamma(\Delta-d/2+1)\Gamma(d/2-m-\ell+2)\over\Gamma(\Delta-m-\ell+3)} 
}}
The integral over $x^+, y^+$ yields
\eqn\intxyp{\eqalign{\!\!\!\!\!\!\int &dx^+ dy^+ 
{  e^{-i {P^2\over 2P^+} (y^+-x^+) } \over (x^++i\epsilon)^{2-\ell-m} (y^+-i\epsilon)^{2-\ell-m} (y^+-x^+-i\epsilon)^{\Delta-d/2+1}  }\longrightarrow_{\ell+m=2}\cr
&={ (2\pi)^2 i^{\Delta-d/2+1}\over \Gamma(\Delta-d/2+1)} \,\, \theta\left({P^+\over 2}\right)\,\theta\left(-{P^2\over 2P^+}\right)\left(-{P^2\over 2 P^+}\right)^{\Delta-d/2-1} \int_0^\infty du u^{\Delta-d/2}\delta^2(1-u)
}}
where we have isolated the divergent term in the dimensionless integral over $u$.

\noindent Gathering all the results, bearing in mind that Eq. \intone\ can be written as
\eqn\vevtwo{\vev{P|T_{--} |P} ={a\over 4} \left(I_{2,0}-2 I_{1,1}+2 I_{0,2}\right) }
leads to
\eqn\vevthree{\vev{P|T_{--} |P} =\tilde{a} \,\,\theta\left({P^+\over 2}\right)\theta\left(-P^2\right)  \,\, \left(-{P^2\over 4}\right)^{\Delta-d/2-1} \left({P^+\over 2}\right)^2}
where
\eqn\coeff{\tilde{a}=-{a\over 4} (2\pi^{d+2})  2 {d-1\over\Gamma(d/2+1) \Gamma(d/2)}
 {1\over \Gamma(\Delta+1)\Gamma(\Delta-d/2+1)} \lim_{\epsilon\rightarrow 0}\delta(\epsilon)}
Given that $a<0$, for any unitary theory where $\Delta-d/2+1\geq 0$, $\tilde{a}$ is positive definite.

\listrefs

\bye